\documentstyle[aps,preprint,eqsecnum,floats,epsf]{revtex}
\tighten
\draft
\begin{document}
\title{An Active-Sterile Neutrino Transformation Solution for
$r$-Process Nucleosynthesis}
\author{G. C. McLaughlin\thanks{Current Address: TRIUMF, 4004
Wesbrook Mall,
    Vancouver, B.C., Canada, V6T2A3.
 Electronic address: {\tt gail@alph01.triumf.ca}}, J.M. 
Fetter\thanks{Current Address: Department of Physics, University of
Wisconsin, Madison, Wisconsin 53706.
Electronic address: {\tt fetter@nucth.physics.wisc.edu}}, A.B.
Balantekin\thanks{Permanent Address: Department of Physics,
University of Wisconsin, Madison, Wisconsin 53706.
Electronic address: {\tt baha@nucth.physics.wisc.edu}}, and G.M.
Fuller\thanks{Permanent
address: Department of Physics, University of California, San Diego
         La Jolla, CA 92093-0319. Electronic address: {\tt
gfuller@ucsd.edu}.}}
\address{ Institute for Nuclear Theory, University of Washington, Box
         351550\\
         Seattle, WA 98195-1550 USA}
\maketitle

\begin{abstract}

  We discuss how matter-enhanced active-sterile neutrino
  transformation in the $\nu_e \rightleftharpoons \nu_s$ and
  $\bar{\nu}_e \rightleftharpoons \bar{\nu}_s$ channels could enable
  the production of the rapid neutron capture ($r$-process) nuclei in
  neutrino-heated supernova ejecta. In this scheme the lightest
  sterile neutrino would be heavier than the $\nu_e$ and split from it
  by a vacuum mass-squared difference of 3 eV$^2 {\ 
    \lower-1.2pt\vbox{\hbox{\rlap{$<$}\lower5pt\vbox{\hbox{$\sim$}}}}\ 
    } \delta m^2_{es} {\ 
    \lower-1.2pt\vbox{\hbox{\rlap{$<$}\lower5pt\vbox{\hbox{$\sim$}}}}\ 
    }$ 70 eV$^2$ with vacuum mixing angle $\sin^2 2\theta_{es} >
  10^{-4}$.

\end{abstract}



\newpage
\section{Introduction}

\indent

In this paper we detail a mechanism through which matter-enhanced
active-sterile neutrino transformation in the $\nu_e
\rightleftharpoons \nu_s$ and $\bar\nu_e \rightleftharpoons
\bar{\nu}_s$ channels could solve the neutron-to-seed nucleus deficit
and alpha-effect problems associated with models of $r$-process
nucleosynthesis from neutrino-heated supernova ejecta. Our solution
makes the production of the $r$-process nuclides in neutrino-heated
supernova ejecta {\it robust} to a wide range of uncertainties in the
neutrino-driven wind models.  Ultimately, our work suggests that heavy
element nucleosynthesis may be a key consideration in constraining the
existence of light sterile neutrinos.

By a sterile neutrino we mean one with interactions which are
significantly weaker than the normal weak interaction. We demand these
interactions to be weak enough so that such a sterile neutrino species
would not contribute appreciably to the decay rate of the $Z^o$
particle.  Our nucleosynthesis considerations are independent of the
details of how the sterile neutrino states are constructed. For
example, many models for sterile neutrinos build these species from
the right-handed Dirac neutrino and left-handed Dirac antineutrino
fields, leaving Majorana active neutrinos $\nu$ and Majorana sterile
neutrinos $N$. In this case $\nu_s$ can be identified with the
left-handed sterile species $N_L$, while $\bar{\nu}_s$ can be
identified with the right-handed sterile neutrino $N_R$.

Of order half of the nuclei with masses $A {\ 
  \lower-1.2pt\vbox{\hbox{\rlap{$>$}\lower5pt\vbox{\hbox{$\sim$}}}}\ 
  }100$ were formed in the rapid neutron capture ($r$-process)
nucleosynthesis scenario\cite{bbfh}. There is as yet no consensus for
the site (or sites) of $r$-process nucleosynthesis, though it seems
likely from meteoritic data-based nucleosynthesis time scale arguments
that one of the sites involves the neutron-rich material associated
with core collapse supernovae \cite{QVW98}. In turn, perhaps the most
compelling model for neutron-rich material ejection following core
collapse supernovae is centered on neutrino heating of material and
the formation of a neutrino-driven \lq\lq wind\rq\rq\ at $\sim
10\,{\rm s}$ after core bounce \cite{neudrw,r1,r2}.

There are, however, a number of difficulties with $r$-process
nucleosynthesis in this model. These difficulties stem principally
from {\it astrophysical} uncertainties in the neutrino-heated outflow
models. In the outflow models, $r$-process nucleosynthesis results
from a freeze-out from nuclear statistical equilibrium.  The
neutrino-heated material is in the form of free nucleons near the
surface of the neutron star, where its neutron-to-proton ratio ($>1$)
is in steady state equilibrium with the $\nu_e$ and $\bar\nu_e$ fluxes
passing through it.  As this material flows out to regions of lower
temperature ($T< 700\,{\rm keV}$) alpha particles are formed, leaving
a sea of free neutrons. Depending on the entropy per baryon, many of
the alpha particles assemble into \lq\lq seed\rq\rq\ nuclei with
masses between $A\approx 50$ and $A\approx100$. As the material flows
further out, to regions of even lower temperature ($T< 300\,{\rm
  keV}$), the free neutrons capture on the seed nuclei to make the
$r$-process nuclear species.

It is clear from this picture that a key quantity for determining the
outcome of the freeze-out process is the neutron to seed nucleus
ratio. It is desirable to have this ratio ${\ 
  \lower-1.2pt\vbox{\hbox{\rlap{$>$}\lower5pt\vbox{\hbox{$\sim$}}}}\ }
100$ in order that the heavier $r$-process species ({\it i.e.}, those
in the $A=195$ peak) can be produced. The neutron to seed nucleus
ratio is determined largely by three quantities: i) the expansion
rate; ii) the neutron-to-proton ratio $n/p$ (or, equivalently, the
electron fraction $Y_e = 1/(1+n/p)$); and iii) the entropy per baryon.
Though different calculations \cite{r1,r2} disagree on the value of
the entropy in the neutrino-driven wind during the r-process
nucleosynthesis, several models can produce values of these three
parameters that yield a high enough neutron-to-seed nucleus ratio at
freeze-out to effect a reasonable $r$-process. Unfortunately there are
neutrino-induced processes operating during or immediately after
freeze-out which can work to greatly reduce the neutron-to-seed
nucleus ratio to the point where acceptable $r$-process
nucleosynthesis in this site would be impossible.  These
neutrino-induced $r$-process destroyers are: i) neutrino neutral
current spallation of alpha particles; and ii) the $\nu_e+n\rightarrow
p+e^-$ reaction accompanying the formation of alpha particles, also
known as the \lq\lq alpha effect.\rq\rq\ 

Meyer pointed out that previously neglected neutrino spallation
reactions on the alpha particles tend to inhibit the $r$-process by
allowing the assembly of too many seed nuclei \cite{meyeralpha}. This
process is especially effective at wrecking the $r$-process where the
entropy is high. A simple steady-state wind model survey of the
thermodynamic parameters in neutrino-heated outflow was conducted by
Qian and Woosley \cite{qw}. These authors concluded that the entropy
in such models should be $\sim100 k$ per baryon, as opposed to Mayle
and Wilson's model with an entropy of $\sim 400 k$ per baryon
\cite{wilson}. In turn, this result might argue against the
effectiveness of neutrino-induced alpha particle spallation in
lowering the neutron to seed nucleus ratio. However, lower entropies
in general imply a lower value of this ratio since there will be more
seed nuclei in these conditions. At best, the neutron to seed nucleus
ratios obtained in lower entropy models are marginal for the
production of the neutron-rich $r$-process species
\cite{putthisin1,meyer97}.

One suggested fix to the low neutron to seed nucleus ratios in these
models is to invoke general relativistic effects \cite{qw,CF}. These
models seem attractive in that they can raise the entropy and increase
the material expansion rate, both of which tend to increase the
neutron to seed nucleus ratio. However, Cardall and Fuller \cite{CF}
found that these models had to be finely tuned in the sense that the
neutron star radius had to be close to that signaling the onset of
dynamical instability. This was required, in turn, in order that the
general relativistic corrections to the outflow rate and entropy be
large enough to solve the neutron to seed nucleus deficit problem.
However, near the dynamical instability radius the differential
gravitational red-shift of $\nu_e$ and $\bar\nu_e$ will act to
increase $Y_e$, partially undoing the beneficial effects of a deeper
gravitational potential well \cite{FQ96}. The general relativistic
fine tuning problem becomes even more extreme if we also demand a
solution to the alpha effect problem.

The alpha effect occurs at the epoch of alpha particle formation. As
the temperature drops, essentially all the protons and most of the
neutrons in the ejecta lock themselves into alpha particles which have
a large binding energy. This phenomenon ultimately will tend to push
the electron fraction higher, towards $Y_e=0.5$. The increase in $Y_e$
comes about because protons produced by electron neutrino capture on
neutrons will in turn capture more neutrons to bind into alpha
particles, reducing the number of free neutrons available for the
$r$-process \cite{alpha}. This effect has been shown to be the biggest
impediment to achieving an acceptable $r$-process yield \cite{MMF}.

One way to avoid or reduce the efficacy of the alpha effect is to
reduce the flux of electron neutrinos at some point above the surface
of the neutron star.  However, in models of the neutrino-driven wind a
large flux of electron neutrinos is required to lift the material off
the surface of the neutron star. In fact since nucleons are
gravitationally bound by about $\sim100\,{\rm MeV}$ near the surface
of the neutron star, and since each neutrino has an energy $\sim
10\,{\rm MeV}$, each nucleon must suffer some $\sim 10$ neutrino
interactions to be ejected to infinity. So if we are to reduce the
$\nu_e$ flux we must do so only at relatively large radius, so that
effective neutrino heating already can have occurred.  Matter-enhanced
neutrino flavor transformation with appropriately chosen difference of
the squares of the masses and vacuum mixing angles can occur above the
heating region yet between the neutrinosphere and where the
$r$-process takes place.

Ordinary active-active neutrino mixing has been extensively studied in
this region of the supernova in the context of rapidly-outflowing
neutrino-heated material\cite{Qetal,QF95,Sigl}. These studies revealed
the important interplay between the material expansion rate and the
$\nu_e$ and $\bar\nu_e$ capture rates on free nucleons. They showed
that a proper understanding of the evolution of the electron fraction
$Y_e$ in response to matter-enhanced neutrino flavor transformation
could not be obtained without due consideration of the expansion rate
and position of fluid elements relative to the neutrino-sphere. The
inherent nonlinearity of the problem demands a coupled treatment for
$Y_e$ and the distribution of neutrino energies.  Early suggestions
that active-sterile neutrino transformation could be important in
supernova dynamics and nucleosynthesis were made with schematic models
and did not include the feedback effects of expansion \cite{Pelt}.

Here we attempt to extend a realistic analysis of neutrino
transformation with coupled outflow to the active-sterile $\nu_e
\rightleftharpoons \nu_s$ and $\bar\nu_e \rightleftharpoons \bar\nu_s$
channels. The interplay of material outflow and active-sterile
neutrino transformation has also been treated for a model with
matter-enhanced active-active and (different) active-sterile channels
by Caldwell, Fuller, and Qian \cite{CFQ98}.

The possibility of a sterile neutrino mixing with an active one was
recently investigated to explain the missing neutrino fluxes in solar,
atmospheric and accelerator neutrino experiments. Recent measurements
of the solar neutrino flux at Superkamiokande \cite{solar} along with
the earlier measurements \cite{othersolar} may indicate mixing of
electron neutrinos with another flavor \cite{solaran}. The measurement
of the atmospheric electron and muon neutrino zenith angle
distributions at Superkamiokande \cite{atm}, taken together with the
lack of observation of $\nu_e$ disappearance at the CHOOZ detector
\cite{chooz} present even a stronger evidence for the mixing of muon
neutrinos with either tau neutrinos or sterile neutrinos \cite{atman}.

Simultaneous interpretation of the solar and atmospheric neutrino
deficits and the $\bar\nu_e$ excess observed by the LSND experiment
\cite{lsnd} in terms of the mixing of only three active neutrinos is
problematic at best \cite{3nus}. These 3-neutrino fits are challenged
by the observed zenith angle dependence of the atmospheric muon
neutrino deficit in Superkamiokande and the establishment of an {\it
  energy dependent} solar neutrino deficit. Indeed, the only
alternatives are to argue that one or more of these neutrino phenomena
is unrelated to neutrino oscillation physics, or to introduce a fourth
neutrino species which, because of the $Z^0$-width limit, must be
sterile. Some time ago it was argued that the LSND data, double beta
decay and cosmological considerations suggested the necessity for
introducing sterile neutrinos \cite{cald}; the recent
experimental/observational data only reinforces these arguments.

In any case, {\it if} there really exist light sterile neutrinos
probably the only way to find out about their properties is to examine
astrophysical environments where neutrinos dominate the dynamics and
nucleosynthesis.  Matter-enhanced active-sterile neutrino
transformation could have a great effect on $r$-process
nucleosynthesis in core-collapse supernovae.  We will show, in fact,
that under the right conditions, such nucleosynthesis is at least as
sensitive as the accelerator experiments to possible mixing of the
sterile and active neutrinos.

In Section II we describe the active-sterile matter-enhancement
process in the post core bounce supernova environment. In Section III
we give a brief description of the supernova outflow and
nucleosynthesis model we employ, while in Section IV we outline our
results. Conclusions are given in Section V.

\section{Matter-Enhanced Transformations of Active to Sterile
Neutrinos}

In the absence of background neutrinos the evolution of flavor
eigenstates in matter is governed by the Schroedinger-like neutrino
amplitude evolution equation \cite{msw}
\begin{equation}
i\hbar \frac{\partial}{\partial r} \left[\begin{array}{cc} \Psi_e(r)
\\ \\ \Psi_s(r) \end{array}\right] = \left[\begin{array}{cc}
\varphi_e(r) & \sqrt{\Lambda} \\ \\ \sqrt{\Lambda} & -\varphi_e(r)
\end{array}\right]
\left[\begin{array}{cc} \Psi_e(r) \\ \\ \Psi_s(r)
  \end{array}\right]\,,
\label{1}
\end{equation}
where
\begin{equation}
  \label{2} \varphi_e(r) = \frac{1}{4 E} \left( \pm
 2 \sqrt{2}\ G_F \left[
  N_e^-(r) - N_e^+(r) - \frac{N_n(r)}{2} \right] E - \delta m^2
  \cos{2\theta_v} \right)
\end{equation}
for the mixing of electron neutrinos (the plus sign on the right-hand
side of the equation) or electron antineutrinos (the minus sign) with
sterile neutrinos.

In these equations
\begin{equation}
  \label{3}
  \sqrt{\Lambda} = \frac{\delta m^2}{4 E}\sin{2\theta_v},
\end{equation}
$\delta m^2 \equiv m_2^2 - m_1^2$ is the vacuum mass-squared
splitting, $\theta_v$ is the vacuum mixing angle, $G_F$ is the Fermi
constant, and $N_e^-(r)$, $N_e^+(r)$, and $N_n(r)$ are the number
density of electrons, positrons, and neutrons respectively in the
medium.  Note that in what follows, we take the sterile neutrino to be
predominantly the heavier mass eigenstate.

We define the potential
\begin{equation}
 \label{v-def}
 V(r) \equiv 2 \sqrt{2} G_F
             \left[N_e^-(r) - N_e^+(r) - \frac{N_n(r)}{2}\right]
\end{equation}
to be proportional to the net weak charge, such that neutrinos
of energy
\begin{equation}
 \label{eres}
 E_{\rm res}(r) \equiv \pm\frac{\delta m^2 \cos 2\theta_v}{V(r)}
\end{equation}
undergo an MSW resonance at a given positive ($\nu_e$) or negative
($\bar\nu_e$) value of the potential.

The mixing of muon and tau neutrinos with sterile neutrinos may be
described similarly.  The evolution Hamiltonian is as for the electron
neutrino species, but with $\varphi_{\mu}$ or $\varphi_{\tau}$
replacing $\varphi_e$ in Eq.~(\ref{1}) as appropriate, where
\begin{equation}
  \label{2a} \varphi_{\mu,\tau}(r) = - \frac{1}{4 E} \left( \pm
  \sqrt{2}\ G_F N_n(r) E + \delta m^2 \cos{2\theta_v}
  \right).
\end{equation}
As before, the $+$ sign corresponds to neutrino mixing, and the $-$
sign to antineutrino mixing.

For a neutral medium we have $Y_p = Y_e$ and $Y_n = 1 - Y_e$, where
$Y_p$ and $Y_n$ give the number of {\it all} protons or neutrons (free
as well as those bound in nuclei), respectively, relative to baryons.
The electron fraction $Y_e$ is given by
\begin{equation}
  \label{4}
Y_e (r)= \frac{N_e^-(r)-N_e^+(r) }{N_e^-(r)-N_e^+(r)+N_n(r)}.
\end{equation}
Inserting Eq.~(\ref{4}) into Eq.~(\ref{2}) one obtains the diagonal
terms in the evolution operator to be
\begin{equation}
  \label{5}
\varphi_e(r) = \pm \frac{3 G_F \rho (r)}{2 \sqrt{2}m_N} \left( Y_e -
  \frac{1}{3} \right) - \frac{\delta m^2}{4E} \cos{2\theta_v},
\end{equation}
and
\begin{equation}
  \label{5a}
\varphi_{\mu,\tau}(r) = \pm \frac{G_F \rho (r)}{2 \sqrt{2}m_N} \left(
Y_e - 1 \right) - \frac{\delta m^2}{4E} \cos{2\theta_v},
\end{equation}
where $\rho(r)$ is the matter density and $m_N$ is the nucleon mass.
Eq.~(\ref{5}) indicates that, with appropriate neutrino parameters and
matter density, for $Y_e > 1/3$ only electron neutrinos and for $Y_e <
1/3$ only electron antineutrinos can undergo an active-sterile MSW
resonance.

The possibility of matter-enhanced conversion of both $\nu_e$'s and
$\bar{\nu}_e$'s can have interesting consequences, but one must
exercise caution. If both electron neutrino and antineutrino fluxes go
through a region of neutrons and protons in equilibrium (i.e. the
reactions $\nu_e + n \rightarrow p + e^-$ and $\bar{\nu}_e + p
\rightarrow n + e^+$ are in steady state equilibrium with the $\nu_e$
and $\bar\nu_e$ fluxes), then no matter what the initial $Y_e$ is one
may naively expect that the system will evolve to a fixed point with
$Y_e = 1/3$.  For example, if initially $Y_e > 1/3$, the $\nu_e$'s
could transform into sterile neutrinos, greatly reducing the rate of
the reaction $\nu_e + n\rightarrow p+ e^-$. However, the rate of
$\bar\nu_e + p \rightarrow n+ e^+$ will remain the same, with the
result that protons will be turned into neutrons, but neutrons will
not be converted back to protons.  Therefore, in this scenario $Y_e$
would be driven lower. If this process brings $Y_e$ below $1/3$, then
$\bar{\nu}_e$'s could be subject to matter-enhanced conversion to
sterile neutrinos.  In this case, the $\bar\nu_e$ flux would be
reduced and so an uncompensated $\nu_e + n\rightarrow p +e^-$ reaction
could push $Y_e$ above $1/3$, and so on.  Indeed an earlier analysis
of the resonant active to sterile conversion was given in Ref.
\cite{nunokawa} where it was argued that the electron fraction becomes
stabilized in supernovae at the fixed point in the evolution,
$Y_e=1/3$, due to the feedback effects. As we will illustrate in the
following sections, our realistic calculations in supernova
neutrino-wind models do not bear out this assessment.

The amount of flavor conversion in an MSW resonance depends on the
adiabaticity of the resonance.  The adiabaticity is a function of the
scale height, $L_V$, of the potential in Eq.  (\ref{v-def}):
\begin{equation}
 \label{v-scaleht}
 L_V = \left|  \left(\frac{1}{\rho}      \frac{d\rho}{dr}\right)
              + \left(\frac{1}{Y_e - 1/3} \frac{dY_e}{dr} \right)
       \right|^{-1}.
\end{equation}
We will refer to the scale height of the potential (or, equivalently,
of weak charges) simply as ``the scale height.''

A pedagogically useful approximation for the neutrino survival
probability is the Landau-Zener approximation \cite{hax-lz,bb-lz},
where the survival probability is directly expressed in terms of the
scale height.  We emphasize that in the results presented in this
paper, exact solutions of Eq.~(\ref{2}) are found numerically and the
Landau-Zener approximation is not employed.

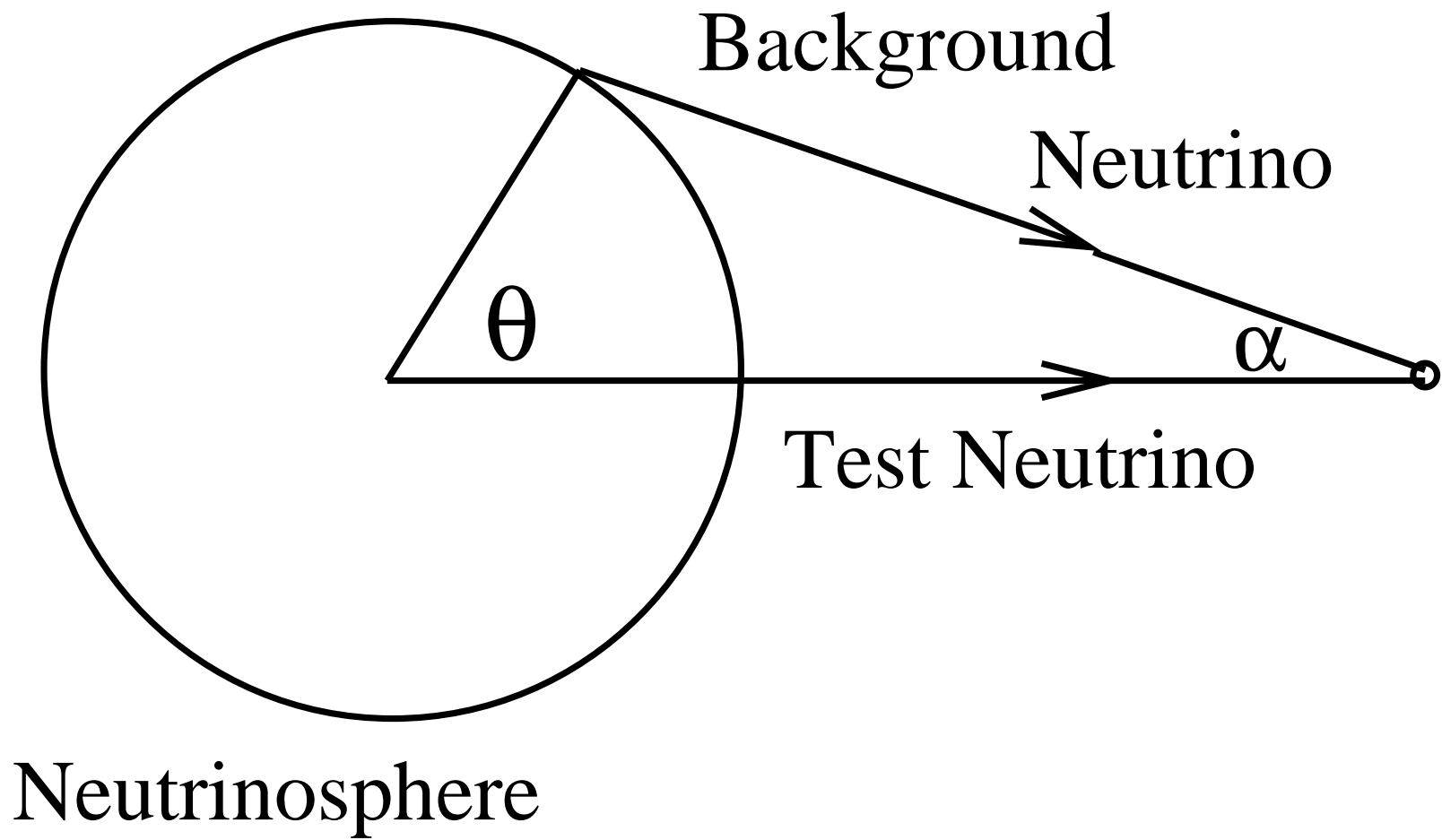
\begin{figure}
\caption{The geometry in the calculation of the effective neutrino
  fraction.}
\label{fig:geometry}
\end{figure}

In the presence of neutrino fluxes (\lq\lq background\rq\rq\ 
neutrinos) the neutrino amplitude evolution Hamilton and the effective
mass in Eq.~(\ref{1}) will have an additional term due to
neutrino-neutrino neutral current forward exchange scattering. In the
case of active-active neutrino evolution, the neutrino background,
because of flavor mixing, contributes both diagonal and off-diagonal
terms in the flavor basis amplitude evolution Hamiltonian,
Eq.~(\ref{1}). However, for active-sterile mixing the off-diagonal
terms are identically zero \cite{sigl}. The diagonal contribution
gives
\begin{equation}
  \label{6}
        \varphi_e(r) = \frac{G_F}{\sqrt{2}} \left[ N_e^-(r) -
  N_e^+(r) - \frac{N_n(r)}{2} + \int d^3{\bf q} \left( 1 - \frac{\bf
  p.q}{|{\bf p}||{\bf q}|} \right) \left( \rho_{\bf q} -
  \overline{\rho}_{\bf q} \right)_{ee} \right] - \frac{\delta
m^2}{4E}
  \cos{2\theta_v},
\end{equation}
where {\bf p} and {\bf q} are the momenta of a test and a background
neutrino respectively, and $(\rho_{\bf q})_{ee}$ is the matrix element
of the single neutrino density matrix operator, $\langle\nu_e|
\hat{\rho} |\nu_e\rangle$.  Here we follow the notation and
terminology of Ref.~\cite{QF95}.  We use the net effective number of
neutrinos per baryon
\begin{equation}
\label{7}
 Y_{\nu} = \frac{m_N}{\rho} \left( N^{\rm eff}_{\nu} - N^{\rm
 eff}_{\bar{\nu}} \right),
\end{equation}
where
\begin{equation}
\label{8}
 N^{\rm eff}_{\nu} \approx \frac{L_{\nu}}{\langle E_\nu \rangle}  
\frac{1}{\pi
R_{\nu}^2}
\int_0^{\infty} dE_{\bf q} \int  d\cos\theta \,
 f_{\nu}(E_{\bf q}, \theta, r) (1 - \cos \alpha),
\end{equation}
where $f_{\nu} (E_{\bf q}, \theta, r)$ is the background neutrino
energy distribution function, which is obtained by evolving forward  
in
time the initial energy distribution function,
\begin{equation}
 \label{dist-evolved}
 f_{\nu} (E_{\nu}, \theta, r)
  = P(E_\nu, \theta, r) f^{\rm initial}_\nu(E_\nu).
\end{equation}
In the above, $L_{\nu}$ is the appropriate neutrino sphere energy
luminosity of the neutrino species in question, $\langle E_\nu
\rangle$ is the average energy of these background neutrinos, and
$P_{\nu} (E_{\bf q},r,\theta)$ is the integrated survival probability
of the background neutrinos. The geometry for neutrino emission is
shown in Figure \ref{fig:geometry} where the angles $\theta$ and
$\alpha$ are defined. In our calculations we actually take the
neutrino energy luminosity to originate at the center of the neutron
star, but we begin the calculations of neutrino amplitude evolution
only above the surface of the star (we take the surface of the neutron
star to be coincident with the neutrino sphere for all species).  This
approximation insures the validity of the overall radial dependence of
neutrino flux in the above expression. It does, however, tend to
underestimate the background neutrino contribution to neutrino
effective mass in the region which is very close to the neutrino
sphere. In practice this is not a problem, as the matter density (the
net electron density) dominates the weak potential in this region; the
density scale height is so small here that neutrino flavor evolution
is non-adiabatic, and so flavor transformation is suppressed.  We have
performed a few calculations with the luminosity originating at a
neutrinosphere, rather than at the center of the neutron star, and
find that our results change very little.  Finally we note that the
survival probability for background neutrinos, $P_{\nu} (E_{\bf
  q},r,\theta)$, in the above expression is \lq\lq short-hand
notation\rq\rq\ for what is in actuality a complicated calculation of
the evolution of the energy distribution functions for background
neutrinos on various trajectories.

With the above definitions Eqs. (\ref{7}) and (\ref{8}) then yield
\begin{equation}
\label{9}
\varphi_e(r) = \pm \frac{G_F \rho (r)}{\sqrt{2} m_N} \left[
\frac{3}{2}\left( Y_e -
  \frac{1}{3} \right) + \left( 2 Y_{\nu_e} + Y_{\nu_{\mu}} +
Y_{\nu_{\tau}} \right) \right]  - \frac{\delta m^2}{4E}
\cos{2\theta_v},
\end{equation}
where, as above, the positive sign in the first term gives the
potential for
$\nu_e$ neutrinos, while the negative sign gives that appropriate for
the
$\bar\nu_e$.

We can see from the above expressions that when $Y_e \approx 1/3$ the
weak potential governing neutrino amplitude evolution will be
dominated by the neutrino background. Wherever the potential is
dominated by the neutrino background the problem of neutrino amplitude
evolution will have an extra degree of nonlinearity. In turn, one
might worry that the problem will become numerically intractable in
this regime. We will argue below, however, that including a realistic
material outflow scheme in this picture facilitates the numerical
calculations and leads to several important features.


\section{Description of the Model}

\subsection{General Features of Outflow}

We consider outflow conditions in the supernova at several seconds
post-core bounce when the $r$-process may take place.  This late epoch
in supernova evolution is easier to treat than the earlier explosion
epoch (shock re-heating epoch) in several respects. It can be hoped
that by this late time, well after the issue of the supernova
explosion is settled, the material above the neutron star will
comprise a very tenuous, nearly hydrostatic envelope.  Indeed, some of
the numerical simulations that can push out this far in time do tend
to bear this out \cite{r1,Qetal,wilsonmayle}. However, convection and
other essentially multi-dimensional phenomena will undoubtedly
operate.  It is not completely understood {\it how} material from near
the surface of the neutron star can be ejected to infinity.  Likely
any such mechanism of ejection will involve neutrino heating, which is
dominated by charged-current reactions. Since each nucleon must
interact with neutrinos $\sim 10$ times in order to be ejected from
the supernova, these neutrino interactions will set $Y_e$ in the
ejecta \cite{Qetal}.  This will be true whether or not the outflow is
one-dimensional in nature or essentially involves convection and
turbulent mixing. Because of this, we can obtain a fair idea of how
$Y_e$ and nucleosynthesis will change under neutrino flavor
transformation by employing a simple one-dimensional outflow picture.

We choose a one dimensional neutrino-driven wind with constant
expansion timescale as our hydrodynamic outflow model.  Additionally,
between the neutron star surface and the radius where the wind
solution is appropriate, we adopt a density gradient and $Y_e$ profile
which gives a good fit to the Mayle and Wilson supernova code results
We also adopt the Mayle and Wilson calculation results for the density
run in the interior of the neutron star, though we ignore feedback
from neutrino physics in this region.  The fitting procedure is
similar to that employed in Ref.  \cite{MFW}. In principle, the
intermediate fit region can noticably change the results, since the
density gradient changes rapidly from very steep at the surface to
very gentle in the wind.  However, we have tested deviations in the
fit and found that they do not change our general conclusions.  Our
scenarios which produce the lowest electron fractions are obtained
with mixing parameters which produce non-adiabatic level crossings in
the steep regions of the density gradient, with the result that
neutrino conversion in such regions is suppressed or inefficient.

In the wind part of our chosen model, the outflow is naturally
homologous: that is, the fluid velocity is proportional to the radial
distance from the neutron star's center, $v \propto r$, where $r$ is
the radial coordinate.  Simple neutrino-driven wind models are
parameterized by two quantities: the assumed constant expansion
timescale $\tau=r/v$, where $r = r_0 \exp(t/\tau)$; and by the entropy
per baryon, $S$.  Additionally we will assume that at sufficiently
large radius the neutrino-heated ejecta will be isentropic, {\it i.e.}
constant entropy per baryon. The relation between the density of an
outgoing fluid element and the time $t$ since it left some initial
point near the neutrino sphere is $\rho \propto \rho_0 \exp(-3 t/
\tau)$, in the wind model.  We adopt this relationship in our
calculations.  Clearly at very large distance from the surface of the
neutron star the exponential acceleration of mass elements will no
longer make sense.  It has been argued, however, that this
approximation will be adequate for the purposes of computing
nucleosynthesis yields in the region above the neutron star before
freeze-out from nuclear statistical equilibrium takes place \cite{qw}.
Our calculations take place prior to this region.

A completely self consistent model to test the effects of neutrino
flavor transformation may not be available for some time.  However,
our results indicate that the neutrino mixing solution for $r$-process
nucleosynthesis is not finely tuned to details of the outflow model.
We therefore choose representative conditions and leave it to the
supernova modeling community to determine, eventually, whether such
outflow and ejection mechanisms can ever be realized.

We assume that the neutrino energy luminosities, initial neutrino
sphere neutrino energy spectral distributions, and the neutron star
radius evolve slowly with time in comparison with the mass outflow
time (time it takes for a mass element which leaves the surface of the
neutron star to finish assembly of alpha particles).  In fact, we can
expect slow evolution of these input quantities over the duration of
roughly one neutrino diffusion timescale at the epoch of $r$-process
nucleosynthesis. Therefore we take the density/temperature gradient,
$Y_e$ profile, and neutrino flavor amplitude distributions with radius
to be fixed throughout our calculations.

We employ $S_{100} = 1$ in our study, since analyses of models of the
neutrino driven wind naturally pick out this entropy scale \cite{qw}.
In what follows we investigate a range of dynamic expansion
timescales, $\tau = 0.1$ s -- $0.9$ s. This range for $\tau$ spans the
regime of plausible wind velocity for supernova models which do not
include an extremely relativistic core.

From the density, entropy and electron fraction, all other
thermodynamic quantities and nuclear statistical equilibrium (NSE)
element abundances may be calculated by taking account of all sources
of entropy in the adiabatically expanding material.  This must be done
without making any assumptions about whether electrons and positrons
are degenerate or relativistic, since they make important
contributions to the entropy.

\subsection{Heuristic Discussion of the Neutrino-Heated Outflow}

Here we give an analytic and heuristic description of neutrino-heated
outflow \cite{qw}. This is meant to serve as a tool for understanding
our numerical results, which, we emphasize, have been calculated
without the use of many of the approximations employed in this
section. One can gain qualitative insight into the general environment
by noting that the neutrino physics and nucleosynthesis are
qualitatively very similar to the process of Big Bang Nucleosynthesis
({\it e.g.}, see Ref.~\cite{Fuller}).

In a wind model at sufficiently large radius (above the heating
regime), the enthalpy per baryon is roughly the gravitational binding
energy of a free baryon, or
\begin{equation}
\label{eq:ts}
T S \approx {\left( { {M_{\rm NS} m_b}\over{m_{\rm
PL}^2}}\right)}{{1}\over{r}}
,
\end{equation}
where $M_{\rm NS} \approx 1.4\,{\rm M_\odot}$ is the mass of the
neutron star, $m_b$ is the mass of a baryon (here we take it to be a
proton), and the Planck mass is defined in terms of Newton's constant
by $m_{\rm PL} \equiv 1/\sqrt{G}$. With these approximations the
radius (in units of ${10}^7\,{\rm cm}$) at which a temperature $T_9
\equiv T/{10}^9\, {\rm K}$ will be found is
\begin{equation}
 r_7 \approx \frac{2.25}{T_9 S_{100}},
\end{equation}
where $S_{100}$ is the entropy per baryon in units of $100$ times
Boltzmann's constant.

In the region above the neutron star where neutrino flavor
transformation can have nucleosynthesis effects, the material will be
radiation dominated and the entropy per baryon will come primarily
from photons and relativistic electron/positron pairs. In this case
\begin{equation}
 S_{100} \approx 3.339 \left(\frac{g_s}{11/2}\right)
 \left(\frac{T_9^3}{\rho_3} \right),
\end{equation}
where $g_s$ is the statistical weight in relativistic particles, and
$\rho_3$ is the rest mass (baryon) density in units of ${10}^3\,{\rm
  g}\,{\rm cm}^{-3}$.  The statistical weight in relativistic
particles will be $g_s \approx 11/2$ where photons and $e^\pm$-pairs
dominate, that is, when $T_9 {\ 
  \lower-1.2pt\vbox{\hbox{\rlap{$>$}\lower5pt\vbox{\hbox{$\sim$}}}}\ }
4$; and $g_s \approx 2$ for $T_9 {\ 
  \lower-1.2pt\vbox{\hbox{\rlap{$<$}\lower5pt\vbox{\hbox{$\sim$}}}}\ }
4$.

An assumed constant entropy per baryon, together with the enthalpy
condition in Eq.~(\ref{eq:ts}), imply that the mass density should
fall off as the inverse cube of the radius. In fact, $\rho \propto
S^{-4} r^{-3}$; for $M_{\rm NS} = 1.4\,{\rm M_\odot}$,
\begin{equation}
 \rho_3 \approx 38 \left(\frac{g_s}{11/2} \right)
 \frac{1}{S_{100}^4 r_7^3}.
\end{equation}
For increasing values of the entropy the density scale height of the
wind-envelope decreases. The density scale height of the baryon
density in the wind will be
\begin{equation}
\label{eqn:dsh}
 L_\rho = {\biggm\vert {d{\ln{\rho}} \over dr}\biggm\vert}^{-1}
\approx
{{1}\over{3}} r \approx L_{\rho 0} {\left( {{M_{\rm
NS}}\over{1.4\,{\rm
M_\odot}}} \right)} {{1}\over{T_9}} {{1}\over{S_{100}}}
\end{equation}
where $L_{\rho 0} \approx 75.0$ km.

The effective scale height of the weak potential ({\it i.e.}, that
relevant for determining neutrino amplitude evolution adiabaticity at
a neutrino mass level crossing) will in general be far more
complicated than that in Eq.~(\ref{eqn:dsh}), since with cumulative
neutrino transformation $Y_e$ will {\it not} be a constant function of
radius and the neutrino background contributions could be large,
especially near neutrino mass level crossings. Just considering the
weak potential stemming from neutrino forward exchange scattering on
the $e^\pm$ component of the plasma, the scale height would be,
\begin{equation}
 \label{eqn:Lweak}
 L_V \approx { \biggm\vert {{3}\over{r}} + {{1}\over{Y_e - 1/3}}
 {d{Y_e} \over dr} \biggm\vert}^{-1} .
\end{equation}
This equation would appear to imply that, for any radius, if $Y_e
\approx 1/3$, then the effective scale height of weak potentials would
be very small, and neutrino and antineutrino evolution through mass
level crossings in these conditions would be non-adiabatic. However,
for $Y_e \approx 1/3$, the weak potential will be dominated by the
neutrino background and this will tend to increase the effective weak
potential scale height and help facilitate adiabaticity.

The expansion rate of the material in the wind, $\lambda_{\rm exp}
\equiv 1/\tau$, is constant when we assume that the dynamic expansion
timescale $\tau$ is constant. We can relate the expansion rate to the
entropy per baryon and the mass outflow rate $\dot{M}=dM/dt$
\cite{Fuller},
\begin{equation}
\label{eqn:lam}
\lambda_{\rm exp} \equiv {{1}\over{\tau}} \approx {\left(
{{45}\over{44\pi^2}}
\right)} {\left( {{11/2}\over{g_s}} \right)} N_A {\left( {{m_{\rm
PL}^2}\over{M_{\rm NS} m_b}} \right)}^3 S^4 \dot{M} .
\end{equation}
Here $N_A=1/m_N$ is Avogadro's number. We note that the expansion rate
is extremely sensitive to the assumed entropy per baryon in the wind.
Expansion timescales $\tau \sim$ a few tenths of a second, and
$S_{100} \sim 1$, will imply mass outflow rates $\dot{M} \sim
{10}^{-6}\,{\rm M_\odot}\,{\rm s^{-1}}$, which are probably adequate
to give an appropriate $r$-process ejection mass per supernova if the
neutron star de-leptonization timescale is long enough.

\subsection{Neutrino Reactions and Initial Neutrino Distribution
Functions}

To model the feedback effect on $Y_e$ from expansion and the
neutrino mixing process we include a numerical calculation of the
reaction rates corresponding to the lepton capture processes:
\begin{equation}
\label{eq:ccf}
\nu_e + {\rm n}  \rightleftharpoons {\rm p}+ e^{-};
\end{equation}
\begin{equation}
\label{eq:ccbf}
\bar{\nu}_e + {\rm p} \rightleftharpoons {\rm n} + e^{+}.
\end{equation}
In the absence of neutrino mixing, the populations of neutrinos and
antineutrinos are comparable, and the forward capture process in Eq.
(\ref{eq:ccf}) is not as fast as that in Eq.~(\ref{eq:ccbf}). This is
a result of the higher average energy of the $\bar\nu_e$ distribution
function (absent significant neutrino flavor transformation).
Ultimately, it is this dominance of the rate for the forward process
in Eq.~(\ref{eq:ccbf}) over the forward process in Eq.~(\ref{eq:ccf})
which produces the neutron-rich conditions which favor the
$r$-process.  Note that the rates at large enough radius of both of
the forward reactions listed above have a $1/r^2$ dependence through
the neutrino flux.

We designate the rates of the reverse processes of electron and
positron capture in Eq.~(\ref{eq:ccf}) and Eq.~(\ref{eq:ccbf}) as
$\lambda_{e^-}$ and $\lambda_{e^+}$, respectively.  At large enough
radius, the forward rates for these processes can be computed as a
function of radius using $\nu_e$ or $\bar\nu_e$ distribution
function-averaged quantities:
\begin{equation}
\label{eq:lamnue}
\lambda_{\nu_e}(r) \approx {\left({{L_{\nu_e}}\over{4 \pi
r^2}}\right)}
{{1}\over{\langle E_{\nu_e}(r)\rangle}}   \langle \sigma_{\nu_e
n}(r)\rangle;
\end{equation}
\begin{equation}
\label{eq:lambar}
\lambda_{\bar\nu_e}(r) \approx {\left({{L_{\bar\nu_e}}\over{4 \pi
r^2}}\right)}
{{1}\over{\langle E_{\bar\nu_e}(r)\rangle}}   \langle
\sigma_{\bar\nu_e
p}(r)\rangle;
\end{equation}
where $\sigma_{\nu_e n}$ and $\sigma_{\bar\nu_e p} $ are the
energy-dependent cross sections for the forward processes in Eq.
(\ref{eq:ccf}) and Eq.~(\ref{eq:ccbf}), respectively, and where angle
brackets represent the appropriate neutrino (or antineutrino)
distribution function averages, as in
\begin{equation}
\label{eq:av}
\langle \sigma_{\nu_e n} (r)\rangle \equiv \int_{\Omega}
{\int_{0}^{\infty}{\sigma_{\nu_e n}(E_\nu) f_{\nu_e} (E_{\nu},
\theta, r) }
dE_\nu} d\Omega_\nu .
\end{equation}
In the absence of neutrino conversion and for large enough radius the
forward rates dominate for both process.  However, the reverse rates
make an important contribution when our calculations begin, even in
the absence of conversion.  After transformation, the forward rate or
rates can be greatly reduced, increasing the importance of the back
reactions.  Therefore we also include positron and electron capture in
our calculations.

The initial neutrino energy spectral distribution functions at the
neutrino sphere are approximated here to be Fermi-Dirac. The
normalized form for these {\it initial}, neutrino sphere energy
distribution functions is then,
\begin{equation}
\label{eq:dist}
f_\nu^{\rm initial}(E_\nu) = {{1}\over{T_\nu^3 F_2(\eta_\nu )}} \cdot
{{E_\nu^2}\over{\exp{(E_\nu/T_\nu -\eta_\nu)} +1}} ,
\end{equation}
where $E_\nu$ is the neutrino (or antineutrino) energy, $\eta_\nu =
\mu_\nu/T_\nu$ is the degeneracy parameter, or neutrino chemical
potential divided by neutrino temperature $T_\nu$. In Eq.
(\ref{eq:dist}) $F_2(\eta_\nu )$ is the relativistic Fermi integral of
rank two ({\it e.g.}, see Ref.~\cite{alpha}, p. 795).

In fact, the true neutrino and antineutrino energy distribution
functions near the neutrino sphere are revealed by detailed transport
calculations \cite{wilsonmayle} to in some cases deviate significantly
from a Fermi-Dirac form, especially on their high energy tails. For
various epochs, degeneracy parameters between $\eta_\nu \approx 0$ and
$\eta_\nu \approx 3$ give the best fits to the numerical results.
Here we will adopt $\eta_\nu =0$ for all initial neutrino energy
distributions.  We take temperatures for our distributions which are
typical of those obtained at 3-10 seconds post bounce, $T_{\nu_e} =
3.5$ MeV; and $T_{\bar\nu_e} = 4.5$ MeV.  Similarly the neutrino and
antineutrino energy luminosities are selected to match (roughly) those
of the numerical transport calculations at this late epoch.  Here we
adopt $L_{\nu_e} =1 \times 10^{51}\,{\rm ergs}\,{\rm s}^{-1}$ and
$L_{\bar\nu_e}=1.3 \times 10^{51}\,{\rm ergs}\,{\rm s}^{-1}$.

In our numerical computations of neutrino flavor mixing, we start with
initial $\nu_e$ and $\bar\nu_e$ distribution functions at the
neutrinosphere in the form given in Eq.~(\ref{eq:dist}), and zero
fluxes of all sterile species. We then evolve the $\nu_e$,
$\bar\nu_e$, $\nu_s$ and $\bar\nu_s$ amplitudes so that the initial
$f_\nu^{\rm initial}(E_\nu)$ evolves into the distribution functions
employed in, for example, Eqs.~(\ref{8}) and (\ref{dist-evolved}):
$f_{\nu} (E_{\nu}, \theta, r) = P(E_\nu, \theta, r) f^{\rm
  initial}_\nu(E_\nu)$.  Note that we perform only radial evolution
($\theta = 0$) to obtain our main results.  Nonradial effects are
discussed at length in Section IV.

\subsection{The Electron Fraction}

The electron fraction is set by competition between the forward
reactions in Eq.~(\ref{eq:ccf}) and Eq.~(\ref{eq:ccbf}) and their
reverse processes of electron and positron capture as outlined above.
The rates for the latter (reverse) reactions, which depend on the
electron temperature and degeneracy parameter decrease much more
rapidly than the rates for the former reactions, which depend on
distance from the neutron star and neutrino distribution function
evolution.  In the limit that the capture rates are very fast in
comparison with the material expansion rate $\lambda_{\rm exp}$ in
Eq.~(\ref{eqn:lam}), and alpha particles are not yet present, the
electron fraction reaches an equilibrium (steady state) value,
\begin{equation}
{\rm Y}_{\rm e} \rightarrow {\rm Y}_{\rm e, eq} \equiv 1/[1 +
(\lambda_{\bar{\nu}_e} + \lambda_{e^-})/(\lambda_{{\nu}_e} +
\lambda_{e^+})].
\label{eq:ye-eq}
\end{equation}

Since $\lambda_{\rm exp}$ is fixed with radius, while both
$\lambda_{\nu_e}$ and $\lambda_{\bar\nu_e}$ fall off with radius,
there will be a point beyond which the lepton capture rates on free
nucleons are very slow in comparison with the expansion rate.  Beyond
this point the electron fraction $Y_e$ will assume a fixed value.  In
our calculations we are not in either the equilibrium (fast neutrino
capture) state or the fixed (no significant capture) state.  In order
to accurately determine the final value of the electron fraction when
alpha particles and neutrino mixing are present, we follow numerically
the evolution of ${ Y}_{ e}$ through the process of weak freeze out.

\subsection{Nucleosynthesis}

Here we review the evolution of a mass element which leaves the
surface of the protoneutron star. Very near the neutron star surface
the material is at quite a high temperature. In fact, the plasma
temperature there will be comparable to the temperatures which
characterize the initial $\nu_e$ and $\bar\nu_e$ distribution
functions.  In this regime NSE will obtain. At an entropy per baryon
$S_{100} \sim 1$, and with the temperature this high, NSE will demand
that the baryons are in free nucleons rather than nuclei, so that only
neutrons and protons are present.  As the material moves out to where
the temperature drops below $T {\ 
  \lower-1.2pt\vbox{\hbox{\rlap{$<$}\lower5pt\vbox{\hbox{$\sim$}}}}\ }
750\,{\rm keV}$, alpha particles begin to form.

As the fluid element moves even further out and the temperature
continues to fall, it becomes energetically favorable to assemble
alpha particles and free neutrons into nuclei - some heavy nuclei
begin to form. At this point the material will begin to freeze out of
NSE through a series of quasi-equilibrium stages \cite{MKC,MMF}.  In
the freeze out from NSE the charged particle reactions fall out of
equilibrium first as a result of the extreme temperature dependence of
their nuclear reaction rates engendered by the large Coulomb barriers
associated with big nuclei. Eventually, the only reactions left in
equilibrium are neutron capture $(n,\gamma)$ and photodisintegration
reactions $(\gamma ,n)$.  This final stage is when neutron capture
builds the heavy nuclei which are the progenitors of the $r$-process
nuclear species we observe in the Galaxy today.

In our calculations, along the trajectory of a fluid element we follow
all thermodynamic and nuclear evolution relevant for $Y_e$ evolution
out to the point at which the first heavy nuclei begin to form.  At
this point $Y_e$ has evolved to where it is essentially fixed,
although a few neutrino capture reactions will still take place which
can alter final element abundance yields \cite{MMF}. The nuclear
equation of state which we employ in our computations is discussed at
length in Ref.~\cite{MFW}.

\subsection{Alpha Effect}

One difficulty in obtaining an adequately low electron fraction $Y_e$
(and, hence, an adequately large neutron to seed nucleus ratio) is the
alpha effect outlined in the introduction.  This effect is the major
impediment to obtaining a successful $r$-process in neutrino-driven
wind models.  As discussed above, as the temperature drops below a
critical value which is dependent on the entropy, alpha particles
begin to form.  Each alpha particle removes two neutrons and two
protons from the free nucleon bath.  Since there were already more
neutrons than protons to begin with, the ratio of free neutrons to
free protons increases.  This would imply that the ratio of free
neutrons to free protons is larger than the $n/p$ ratio characteristic
of weak steady state equilibrium.  However, on a fairly rapid
timescale (just how rapid depends on the magnitudes of
$\lambda_{\nu_e}$ and $\lambda_{\bar\nu_e}$), neutrino captures will
occur on these free neutrons, turning them into protons.  Some protons
will turn to neutrons as well, but the overall ratio of free neutrons
to free protons will decrease, and, hence, the overall $Y_e$ will
increase. The ultimate result of this process is that the neutron to
seed nucleus ratio will decrease from what it would have been had
alpha particles not formed.  This is the alpha effect identified in
Ref.~\cite{alpha}.

In the alpha effect there are no compensating $\nu_e$ or $\bar\nu_e$
captures on alpha particles, since for the expected electron neutrino
and electron antineutrino energy spectra, alpha particles are
basically inert with respect to the charged current interactions.
However, neutrinos and antineutrinos can still capture on nuclei and
$\nu_e + n \rightarrow p +e^-$ can continue well into the neutron
capture region (between $ 3{\ 
  \lower-1.2pt\vbox{\hbox{\rlap{$>$}\lower5pt\vbox{\hbox{$\sim$}}}}\ }
T_9 {\ 
  \lower-1.2pt\vbox{\hbox{\rlap{$>$}\lower5pt\vbox{\hbox{$\sim$}}}}\ }
1$), further robbing the $r$-process of its requisite neutrons
\cite{MMF}.

Our numerical computations include a treatment of alpha particle and
heavy \lq\lq seed\rq\rq\ nucleus formation as in Ref.~\cite{MFW}.
Since we also include numerical integration of the rates in Eq.
(\ref{eq:lamnue}) and Eq.~(\ref{eq:lambar}), our calculations will
follow accurately the run-up of $Y_e$ resulting from the alpha effect.
As we will show, active-sterile neutrino transformation can result in
a depressed $\nu_e$ flux relative to the no-neutrino oscillation
scenario and, in turn, this can suppress the alpha effect.

\section{Discussion of Results}

We cast our results in terms of electron fraction $Y_e$ at the time of
the formation of heavy nuclei at $T_9 \approx 3$. This corresponds to
a radius where the weak capture rates have become quite slow.
Computing the evolution of the system out this far allows us to
faithfully follow any potential alpha effect.  Throughout the
calculation, we perform a full numerical solution of the MSW evolution
equations, neglecting only the effects of the neutrino background.  No
approximation is employed, so we can track the detailed behavior of
the neutrinos as they pass through their resonances.  In understanding
the results, however, it will be useful to keep in mind the dependence
of survival probability on scale height.  In our scenario, where
neutrino evolution begins at very large potentials and the vacuum
mixing angle is small, passage through an adiabatic resonance (with a
large scale height) gives near-complete conversion; passage through a
non-adiabatic resonance (with a small scale height) yields very little
conversion.

We couple the MSW evolution to a numerical calculation which self
consistently determines temperature, electron chemical potential and
nuclear statistical equilibrium abundances of protons, neutrons, and
alpha particles, from the entropy, density and electron fraction at
each time step.  Also at each time step, all weak capture rates are
calculated, using the survival probabilities for each neutrino and
antineutrino energy bin, and the electron fraction $Y_e$ is updated.
This new $Y_e$ is then used to determine the new neutrino survival
probabilities, as well as the updated thermodynamic and abundance
variables. Implementation of this feedback effect is the substantially
new element in our approach to neutrino transformations in supernovae.
At very high densities, the electron fraction is set primarily by
degeneracy.  Its increase before $\rho = 10^8$ g/cm$^3$ is controlled
initially by neutrino capture on neutrons, Eq.~(\ref{eq:ccbf}).  As
the degeneracy is lifted, positron capture on neutrons further raises
and contributes quite significantly to the electron fraction.  Thus,
because of the importance of positron and electron capture at small
radius, we expect feedback effects to be small in the region where we
use a static profile of $Y_e$.  For the highest densities ($\rho > 4
\times 10^8$ g/cm$^3$), then, we use a static electron fraction
profile to compute the MSW evolution of the neutrino amplitudes.

\subsection{The Mechanism}

In this section we describe the numerical evolution of neutrinos and
antineutrinos in concert with the composition and the $Y_e$ value of
outflowing mass elements.  We focus on the feedback mechanism: roughly
(neglecting the neutrino background terms) the value of $Y_e$
determines whether neutrino flavor transformation takes place, while
$\nu_e$ and $\bar\nu_e$ captures determine $Y_e$.

Here we do not follow the evolution of the $\nu_\mu$, $\bar\nu_\mu$,
$\nu_\tau$, and $\bar\nu_\tau$ distribution functions. However,
depending on the adopted neutrino mass level schemes, there could well
be transformations either among these species or with $\nu_e$'s, or
even with sterile neutrino species. The possible effects of some of
these types of neutrino transformation channels are treated elsewhere
\cite{CFQ98}. In principle neutrino transformation among the mu and
tau neutrinos and sterile species could affect the neutrino background
which is partly responsible for driving $\nu_e$ and $\bar\nu_e$
evolution.

We find that neutrino background effects are everywhere sub-dominant,
except very close to where $Y_e = 1/3$.  The neutrino background in
principle can change both the position of the resonance and the scale
height.  (Recall that ``the scale height'' is the scale height of weak
potential.)  However, since the density gradient is so steep, the
small contribution from the background has little impact on the
resonance position.

For large neutrino energies, the scale height at resonance is
dominated by the derivative of $Y_e$ near $1/3$ before we introduce
background effects.  The neutrino background may significantly change
the conversion probability for those neutrinos.  This change affects
neutrinos of much higher energy than the ones we consider, and so we
neglect the background in the following discussion. For example, we
estimate that background makes an unimportant contribution to the
scale height for neutrinos below 50 MeV as long as $\delta m^2 \gtrsim
0.1$.  We will consider the effects of including neutrino background
in more detail after the discussion of the main result.

The potential for electron neutrino and electron antineutrino
active-sterile transformation is controlled by density and electron
fraction, since the potential $V \propto \rho (Y_e - 1/3)$.  When the
potential is positive, the $\nu_e \rightleftharpoons \nu_s$ channel
operates, and when the potential is negative, the $\bar\nu_e
\rightleftharpoons \bar\nu_s$ transformation can take place.

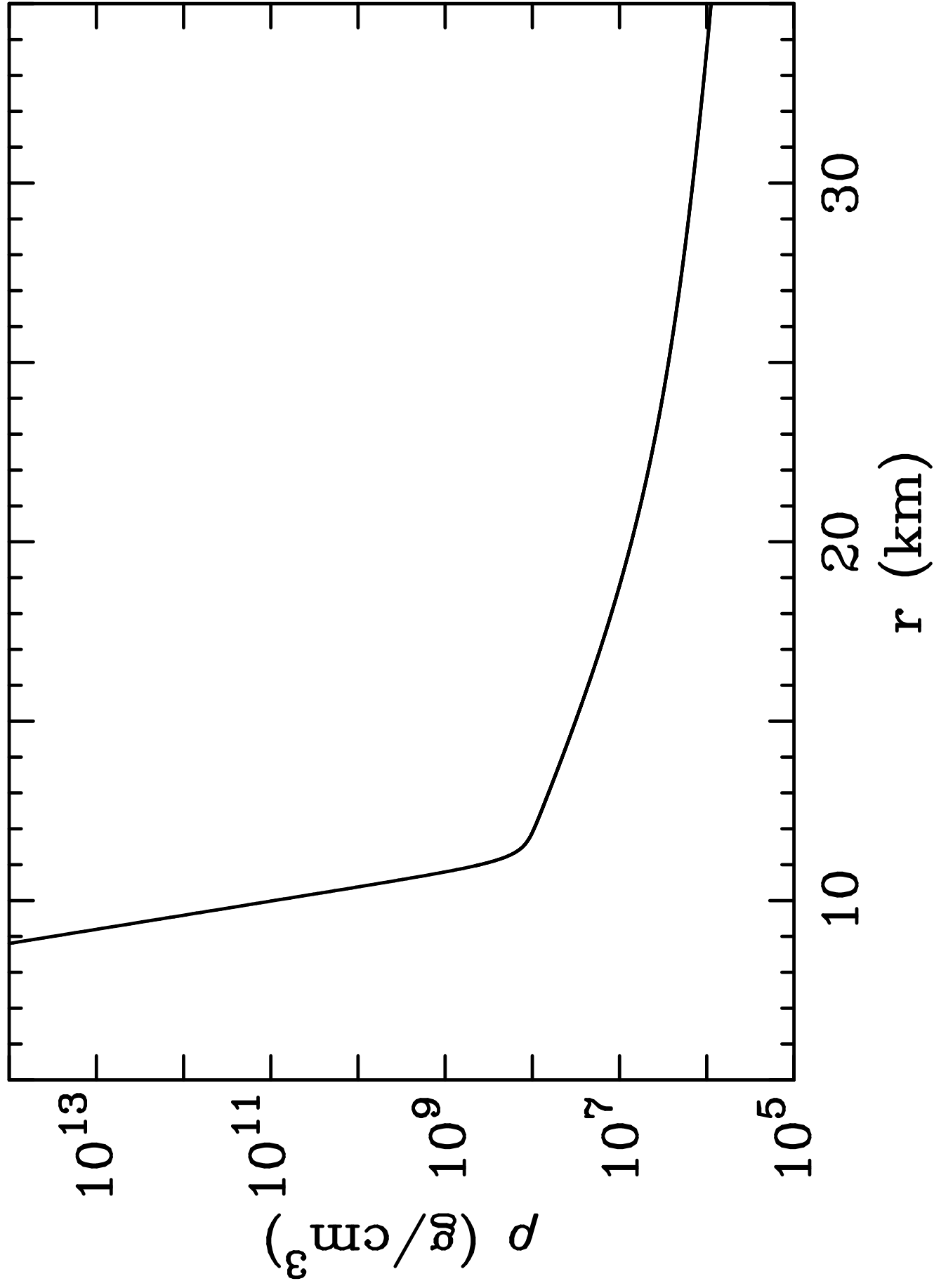
\begin{figure}
\caption{ Density is plotted against distance as measured from the
center of the neutron star.}
\label{fig:density}
\end{figure}

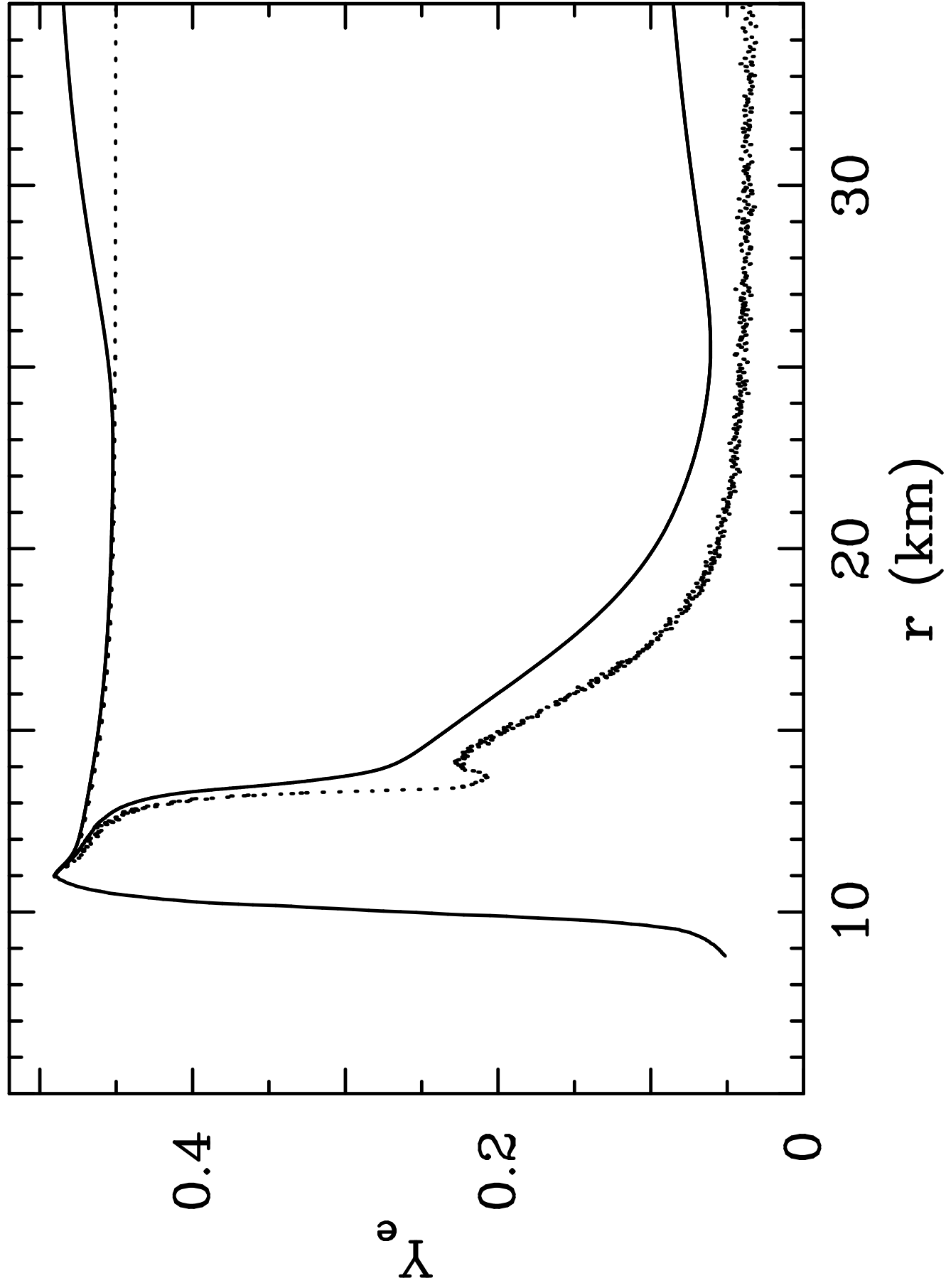
\begin{figure}

\caption{The upper pair of curves shows the actual and equilibrium
electron fraction in the absence of any flavor transformation.  The
lower pair of curves shows the same with neutrino mixing parameters as
in Figures \ref{fig:rates} and \ref{fig:energy}.  In each pair, the
lower line corresponds to the equilibrium ${\rm Y}_{\rm e}$.  Above 11
km we include the effects of feedback.  Above ${\rm Y}_{\rm e} = 1/3$,
electron neutrinos may undergo flavor transformation, while below
${\rm Y}_{\rm e} = 1/3$ electron antineutrinos may transform. The
neutrino driven wind parameters are the same as in Figures
\ref{fig:energy} and \ref{fig:rates}.  For this dynamical timescale,
the actual ${\rm Y}_{\rm e}$ closely tracks the equilibrium ${\rm
Y}_{\rm e}$.  The near complete transformation of electron neutrinos
drives the electron fraction to very low values in the lower set of
curves.  In addition, it almost completely suppresses the alpha
effect.}
\label{fig:ye}
\end{figure}

At the surface of the protoneutron star the density profile is very
steeply falling, while the electron fraction $Y_e$ is rising, as can
be seen in Figures \ref{fig:density} and \ref{fig:ye}.  Since $Y_e <
1/3$ very near the neutron star surface, there is an initial resonance
for $\bar\nu_e \rightleftharpoons \bar\nu_s$ at high density, $\rho =
2-3 \, \times \, 10^9 {\rm \,g\, cm}^{-3}$.  Immediately following
this antineutrino resonance is an electron neutrino resonance ($\nu_e
\rightleftharpoons \nu_s$), as $Y_e$ passes above 1/3.  In this
region, the density profile is steep; $Y_e$ is changing rapidly with
radius; and $Y_e \approx 1/3$.  Therefore, the scale height is tiny
(see Eq.~\ref{v-scaleht}), so these resonances are usually quite
nonadiabatic and do not yield significant flavor transformation.  At
very large $\delta m^2 \sin^2 2 \theta$, the resonances may become
adiabatic.  We will return to this point below.

For regions above the protoneutron star where $T_9 {\ 
  \lower-1.2pt\vbox{\hbox{\rlap{$<$}\lower5pt\vbox{\hbox{$\sim$}}}}\ }
25$, the outflow goes over to the neutrino-driven wind solution with
constant dynamic expansion timescale, and we begin to include feedback
effects in our calculation.  In the wind the density continues to fall
with increasing distance, although much less steeply than at the
surface as can be seen from Eq.~(\ref{eqn:dsh}).  In this region $Y_e$
roughly levels off with radius (before neutrino mixing effects), and
the falling density allows the electron neutrinos to pass through a
resonance.  The density gradient is much smaller here, and so this
resonance is likely to result in more adiabatic flavor transformation,
for a wide range of neutrino mixing parameters and for a broad range
of neutrino energies.

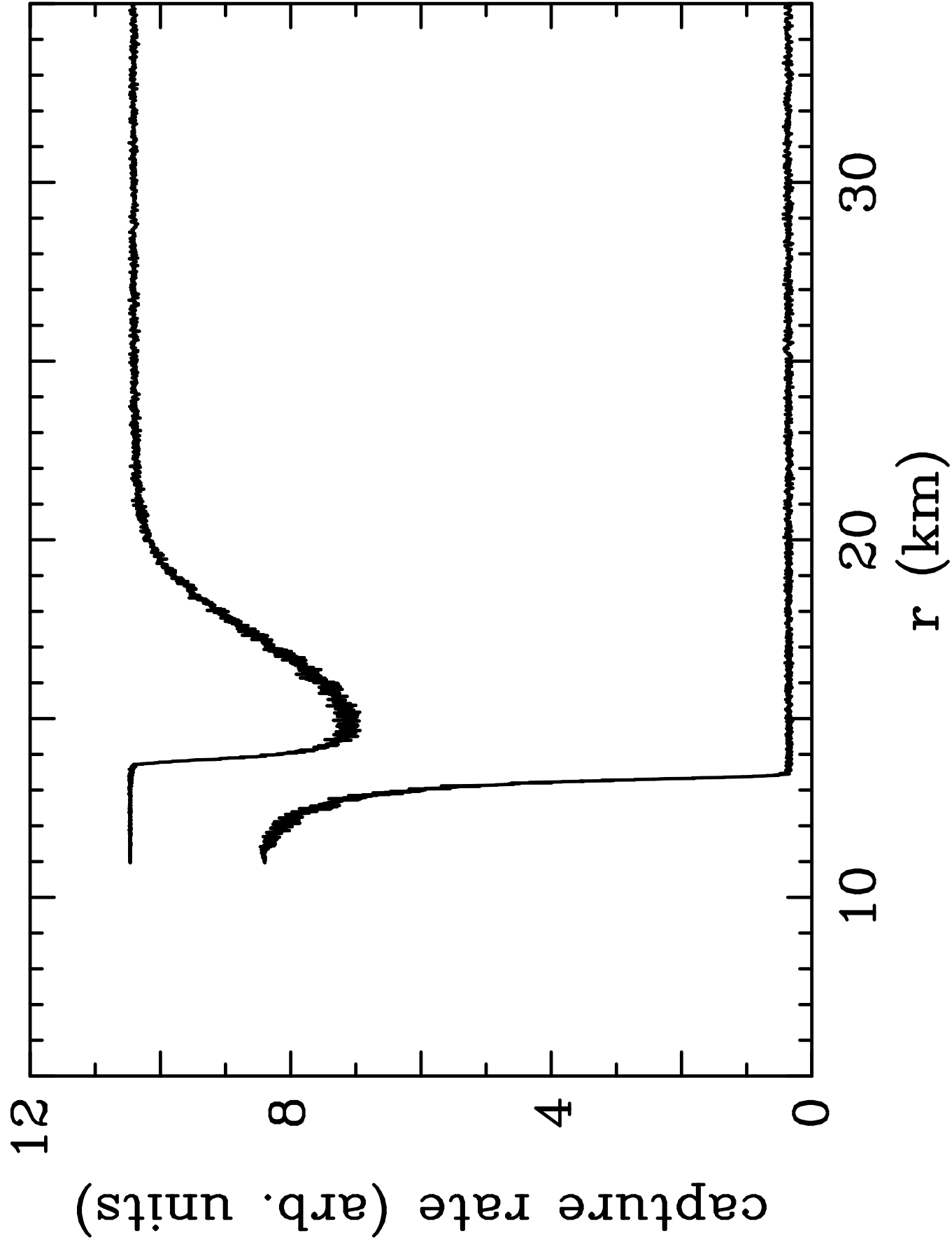
\begin{figure}
\caption{Electron neutrino (lower curve) and electron antineutrino
(upper curve) capture rates on neutrons and protons respectively,
plotted against $r$, the distance from the center of the protoneutron
star, for the same choice of parameters as in Figure \ref{fig:energy}.
The $1/r^2$ dependence of the neutrino flux has been removed for
illustrative purposes only. All variation seen in the capture rates is
due to transformation into sterile neutrinos.}
\label{fig:rates}
\end{figure}

This transformation is visible in Figure \ref{fig:rates}, where we
show the electron neutrino capture rate on neutrons as a function of
radius (with the $1/r^2$ dependence of the neutrino flux divided out).
Since we begin at high density, and therefore large weak potential,
low-energy $\nu_e$'s transform to sterile neutrinos first.  As the
potential falls with density, higher energy $\nu_e$'s transform.

For a sufficiently long dynamical expansion timescale (small
$\lambda_{\rm exp}$) as in this example, the plummeting $\nu_e$
capture rate, $\lambda_{\nu_e}$, eventually falls well below
$\lambda_{\bar\nu_e}$.  This unbalances the weak steady state
equilibrium and tends to shift it in favor of the reaction $\bar\nu_e
+ p \rightarrow n +e^+$. Therefore $Y_e$ is driven down, as can be
seen in Figure \ref{fig:ye}.  This figure shows both the equilibrium
electron fraction (Eq.~\ref{eq:ye-eq}) and the actual electron
fraction.  The local minimum in the equilibrium electron fraction
occurs when the $\nu_e$'s disappear.  The equilibrium $Y_e$ is not
driven all the way to zero at this point, since positron capture on
neutrons is still marginally significant.

A decreasing $Y_e$ causes $V(r)$ to decrease more quickly than would
be the case were $Y_e$ to remain fixed. This behavior can be seen in
Figure \ref{fig:potl}. Another consequence of the rapid decrease in
$Y_e$ is that the scale height becomes smaller (see
Eq.~\ref{eqn:Lweak}) and neutrino amplitude evolution through the
$\nu_e \rightleftharpoons \nu_s$ resonances becomes somewhat less
adiabatic.  Conversion of high-energy electron neutrinos, then, is
slightly less efficient than conversion of low-energy neutrinos.
However, for a large range of neutrino mixing parameters, almost all
electron neutrinos transform into sterile states.

More importantly, the disappearance of the $\nu_e$'s can push the
electron fraction to values $Y_e < 1/3$, making $V$ negative.  As a
result, high energy $\bar\nu_e$'s will undergo a resonance and convert
to $\bar\nu_s$'s.  This does not usually drive $Y_e$ back up, since
there are so few $\nu_e$'s left.  However, it slows down the fall of
$Y_e$ with radius, creating a \lq\lq knee\rq\rq\ in the actual
electron fraction curve at around 14 km in Figure \ref{fig:ye}.

The beginning of the return of the electron antineutrinos is marked by
the local maximum in $Y_e$.  When $Y_e$ is near 1/3, the change in the
weak potential is dominated by the change in the electron fraction.
However, when $Y_e$ is far from 1/3, the change in the potential is
dominated by the change in the density, which is falling rapidly.
This tends to pull the potential toward zero, turning it over and
preventing lower-energy $\bar{\nu}_e$'s from transforming and causing
the higher energy $\bar\nu_s$'s, which resulted from $\bar\nu_e
\rightarrow \bar\nu_s$, to transform back to active states.  This
regeneration of $\bar\nu_e$'s results in the recovery of nearly their
full population (Figure \ref{fig:rates}).  Along with the decreasing
importance of positron capture, the regeneration of the $\bar\nu_e$'s
allows the electron fraction to fall to very low values.  Figure
\ref{fig:energy} shows explicitly the energy of the $\nu_e$'s and
$\bar{\nu}_e$'s which undergo a resonance at a given position.

In our calculations, the choice of the dynamic expansion timescale
plays a crucial role in determining the final electron fraction.  With
a relatively long timescale (as in our example), many neutrino and
antineutrino captures are possible, and the actual $Y_e$ closely
tracks the equilibrium $Y_e$.  A longer dynamic expansion timescale
also augments the alpha effect if many $\nu_e$'s are still present.
In Figure \ref{fig:ye}, a small alpha effect can be seen as the upturn
in $Y_e$ at around 25 km.

\subsection{Variation of Parameters}

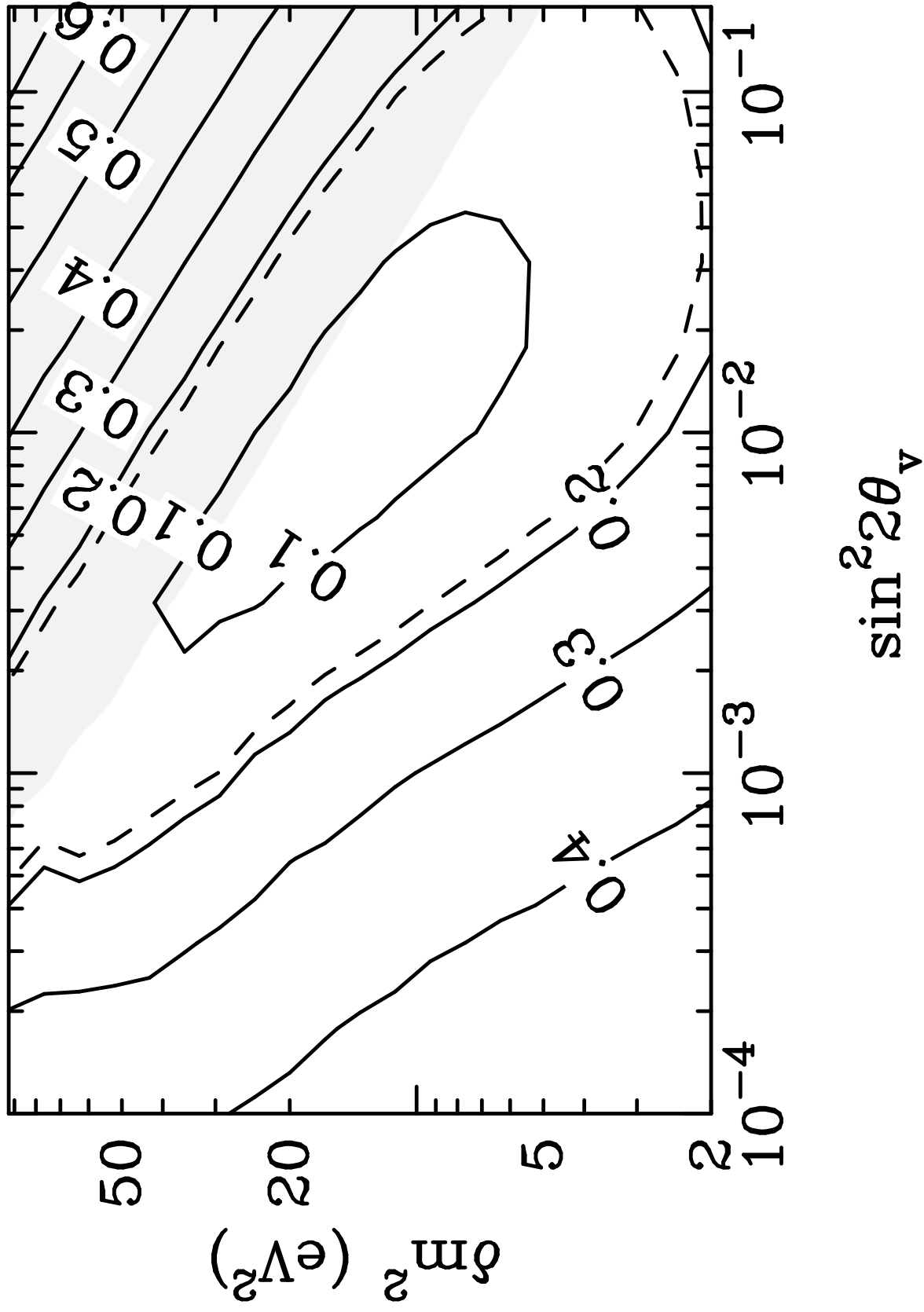
\begin{figure}
\epsfxsize=14cm
\epsfxsize=14cm
\caption{Contours of electron fraction at the time of heavy (${\rm A} >
  40$) element formation, for a range of neutrino mixing parameters
  $\delta m^2$ and $\sin^2 2 \theta_{\rm v}$.  The neutrino driven
  wind timescale is 0.3 seconds.  The conditions necessary for a
  neutron-to-seed ratio of at least 100 are within the ${\rm Y}_{\rm
    e} = 0.18$ (dashed) contour. If no flavor transformation takes
  place, ${\rm Y}_{\rm e} = 0.49$. In the gray region nonradial
  neutrino paths (not included in this example) may be significant.}
\label{fig:cont1}
\end{figure}

Figure \ref{fig:cont1} explores the effect of variation in the
neutrino mixing parameters.  Here we plot the value of $Y_e$, just as
alpha particle formation is ending and heavy nucleus formation is
beginning, for a wind model with a dynamical expansion timescale of
$\tau = 0.3$ s.  One indication of whether $r$-process nucleosynthesis
may successfully occur is if the neutron to seed nucleus ratio ($R$)
at the time rapid neutron capture begins is around 100.  If $R$ is
considerably smaller than 100, then the A=195 peak will not form.
According to Meyer and Brown \cite{meyer97}, for a dynamic expansion
timescale of 0.3 seconds $R> 100$ is possible if $Y_e < 0.18$. (Note
that the definition of our timescales differs; ours is three times
theirs.)  The $Y_e = 0.18$ contour is shown as the dotted line in
Figure \ref{fig:cont1}.  As mentioned above, there can be further
change in the electron fraction during the first stages of heavy
nucleus formation (after our calculation ends).  However, if the
electron neutrino survival probability is small, we expect this change
to be minimal.

The electron fraction is larger than is desirable for $r$-process
nucleosynthesis on both the upper right and lower left sides of the
figure.  We compare these regions to the optimal behavior (center of
the figure) which was described in detail above.  As $\delta m^2
\sin^2 2 \theta$ decreases, conversion of electron neutrinos is less
effective, since evolution through the resonances becomes less
adiabatic.  In the lower left corner of the plot, it can be seen that
$Y_e$ is asymptotically approaching the value it takes on without
neutrino mixing.

As $\delta m^2 \sin^2 2 \theta$ increases, it becomes possible to have
flavor transformation proceed through the resonances which are closest
to the neutron star surface. This can result in some of the electron
neutrinos and antineutrinos being partially converted to sterile
species before they leave the vicinity of the surface of the
protoneutron star.  When the $\nu_e$'s encounter the later resonances,
those that were previously converted to steriles can convert back to
electron neutrinos.  This leaves a partial complement of electron
neutrinos which causes $Y_e$ to drop less than in our optimal example.

When $\delta m^2 \sin^2 2 \theta$ is large, most $\nu_e$'s and
$\bar{\nu}_e$'s convert to sterile states in the inner resonances.
Then, in the wind region, $\nu_s$'s convert back to active states and
the $\nu_e$'s convert to sterile states.  For sufficiently large
$\delta m^2 \sin^2 2 \theta$, we end up with a large population of
active $\nu_e$'s and a smaller population of $\bar{\nu}_e$'s.  This
drives the electron fraction above 0.5 and the later $\bar{\nu}_e$
mass level crossings never occur.

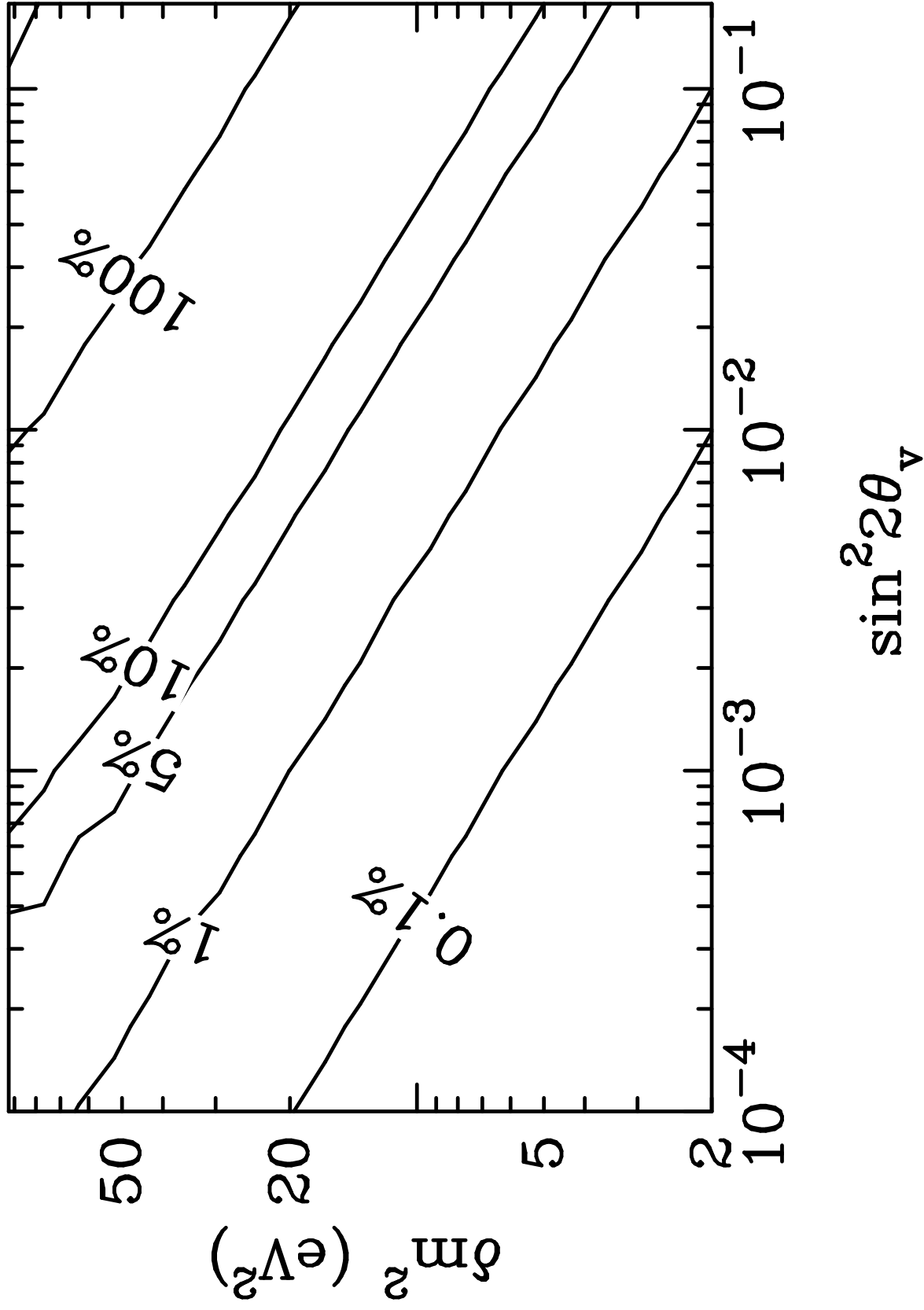
\begin{figure}
\caption{The percentage difference in the $\nu_e$ capture rate at 11
  km, between calculations including and not including the effects of
  nonradial neutrino paths.}
\label{fig:nonradial}
\end{figure}

In this region nonradial neutrino paths play an important role in the
neutrino evolution.  Neutrinos which leave the neutrino sphere
nonradially find the inner resonances more adiabatic, since they
encounter them at a grazing angle. Therefore, if we consider nonradial
neutrino paths, the inmost resonance will begin to cause
transformation at lower $\delta m^2 \sin^2 2 \theta$.  We do not
include these effects in Figure \ref{fig:cont1}, but we estimate their
importance by computing the relative difference in the $\nu_e$ capture
rate on neutrons at 11 km with and without nonradial effects:
\begin{equation}
 \label{pctdiff}
 {\rm relative\ difference} = \frac{\rm(radial\ rate) -
(nonradial\ rate)}
                                   {\rm (nonradial\ rate)}.
\end{equation}
(As can be guessed readily from the neutrino emission geometry,
nonradial neutrino paths have very little effect at larger radii.)
The relative difference is shown in Figure \ref{fig:nonradial}. The
dotted line in Figure \ref{fig:cont1} corresponds to a 10\% reduction
in the $\nu_e$ capture rate from nonradial effects.  Above this line,
a full treatment of neutrino oscillations in the presence of neutrino
scattering at high density, including nonradial effects would be
necessary in order to fully understand the implications of these
neutrino mixing parameters for $r$-process nucleosynthesis.

Although the lines of constant $\delta m^2 \sin^2 2 \theta$ describe
much of the behavior seen in the plot, there is additional variation
above and below the island of lowest electron fraction. As $\delta
m^2$ decreases along a line of constant $\delta m^2 \sin^2 2 \theta$,
the density at which the electron neutrino resonance occurs decreases
and the distance from the protoneutron star increases.  At larger
distance the neutrino capture rates are smaller, due to the $1/r^2$
dependence in the neutrino fluxes, so the actual $Y_e$ approaches the
equilibrium $Y_e$ more slowly. Therefore, the final $Y_e$ is higher.
On the other hand, as $\delta m^2$ increases, the electron neutrinos
convert at higher density, where the scale height is smaller, and
therefore $\nu_e$ conversion is less efficient. The $\nu_e$'s which
survive conversion in this case cause a larger alpha effect.

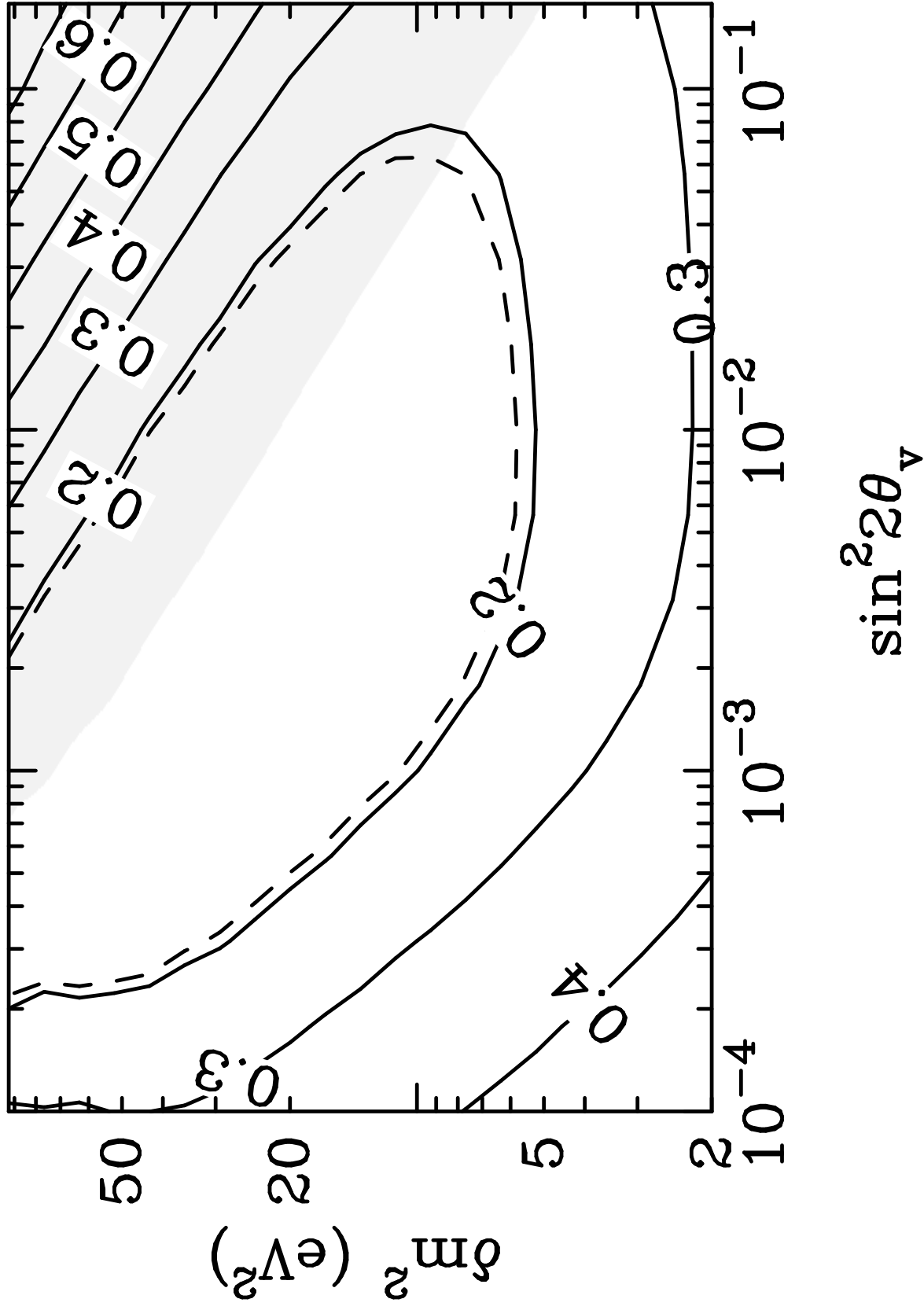
\begin{figure}
\caption{As in Figure \ref{fig:cont1}, but for a timescale of 0.1
  seconds. The conditions necessary for a neutron to seed ratio of at
  least 100 are within the ${\rm Y}_{\rm e} = 0.19$ (dashed) contour.
  If no flavor transformation takes place, ${\rm Y}_{\rm e} = 0.47$.}
\label{fig:cont2}
\end{figure}

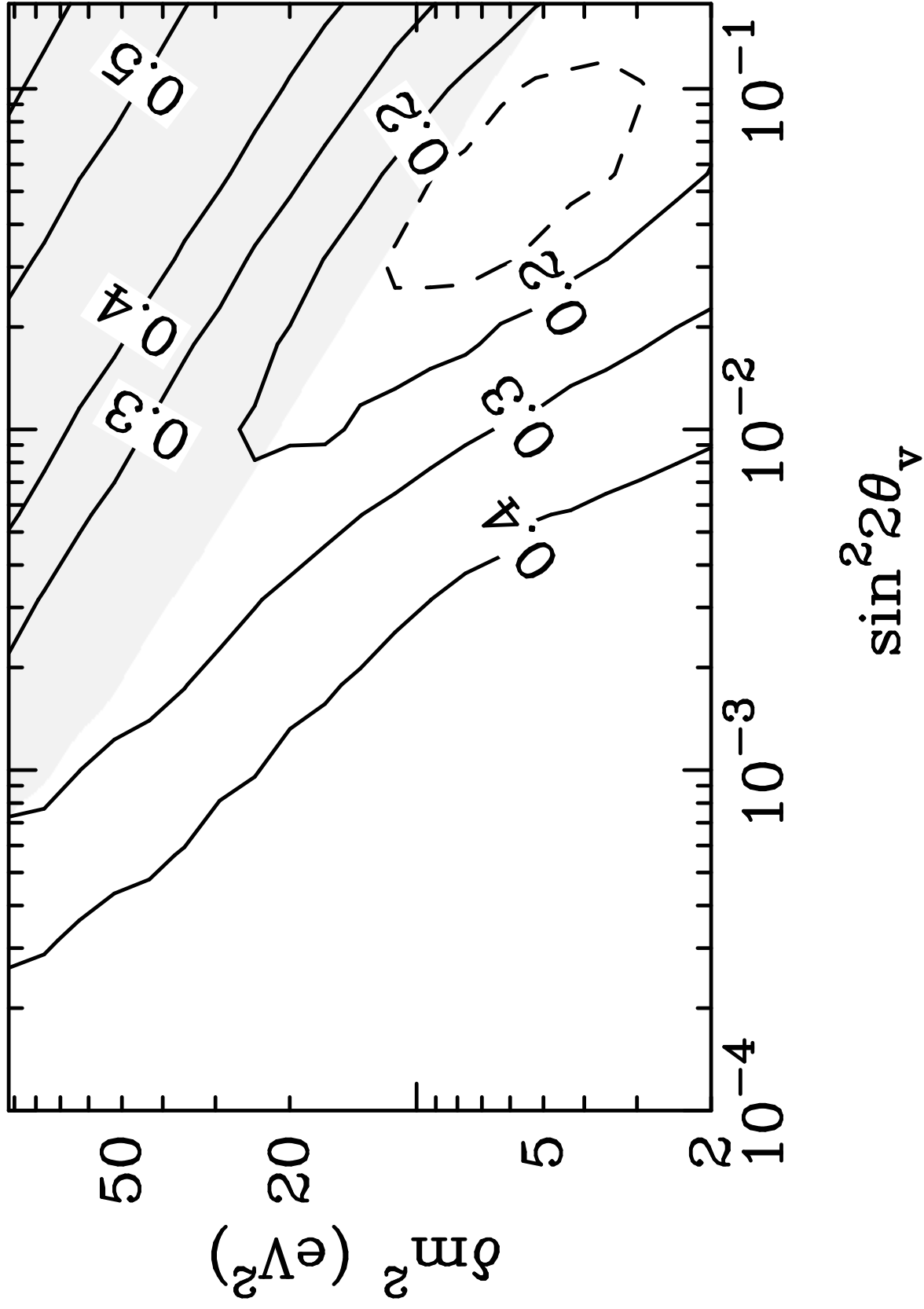
\begin{figure}
\caption{As in Figure \ref{fig:cont1}, but for a timescale of 0.9
  seconds. The conditions necessary for a neutron-to-seed ratio of 100
  are within the ${\rm Y}_{\rm e} = 0.15$ (dashed) contour. If no
  flavor transformation takes place, ${\rm Y}_{\rm e} = 0.50$.}
\label{fig:cont3}
\end{figure}

In Figures \ref{fig:cont2} and \ref{fig:cont3}, we investigate the
impact of varying the dynamic expansion timescale.  With a longer
timescale (Figure \ref{fig:cont3}), there is more time for the actual
$Y_e$ to approach the equilibrium value, but also a stronger alpha
effect.  The alpha effect pulls the final electron fraction toward
$0.5$, shifting all the contours away from the $0.5$ contour, as
compared with the shorter expansion timescale case.

We will now consider the movement of the island of smallest $Y_e$ as
the dynamic expansion timescale varies.  The island moves up and down
in $\delta m^2$, roughly along lines of constant $\delta m^2 \sin^2
2\theta$. There is a different location for the optimal island of
parameter space because different resonance locations are optimal in
reducing $Y_e$ for different dynamic expansion timescales.

There are three ways that the resonance location affects the final
value of $Y_e$.  First, the closer the neutrino mass level crossing
position is to the surface of the protoneutron star, the more readily
the actual value of $Y_e$ will track the equilibrium value. This is
mostly because of the $1/r^2$ dependence of the neutrino flux and the
associated radial dependence of $\lambda_{\nu_e}$ and
$\lambda_{\bar\nu_e}$.  Second, the resonance position---and thus its
adiabaticity---affects the number of $\nu_e$'s present when alpha
particles form, and therefore partly determines the strength of the
alpha effect.  Finally, the two $\bar{\nu}_e$ resonances in the wind
region may have different adiabaticities.

For the very short ranges of dynamic expansion timescale, the optimal
island in parameter space moves to larger values of $\delta m^2$,
where the resonances are closer in general to the surface of the
neutron star.  A deep resonance produces the lowest $Y_e$ at short
expansion timescale primarily because it is necessary to convert the
$\nu_e$'s where the neutrino fluxes are large, so that the actual
value of $Y_e$ tracks its equilibrium value.  At short expansion
timescale the electron fraction falls out of weak equilibrium quickly.
Since a deep resonance is less adiabatic, it leaves a larger
population of residual $\nu_e$'s when alpha particles form.  However,
at short dynamic expansion timescale, the alpha effect can be quite
small, so that the surviving $\nu_e$'s have little effect.  Since the
electron fraction freezes out so quickly at short dynamic expansion
timescale (the weak freeze out radius is small in this case), the
population of $\bar{\nu}_e$'s becomes relatively unimportant by the
time they begin to convert to sterile species.

Conversely, at long dynamic expansion timescale, the island of optimal
parameter space for reducing $Y_e$ moves to smaller values of $\delta
m^2$.  The resonances in this case are farther from the surface of the
neutron star.  Here it is not so important to convert the $\nu_e$'s
deep in the supernova, because in this scenario there will be time for
them to affect the electron fraction even if their flux is relatively
small.

It is, however, important to convert as many $\nu_e$'s as possible in
this case, if the pernicious increase in $Y_e$ stemming from the alpha
effect is to be minimized.  Finally, for this scenario, the population
of $\bar{\nu}_e$'s at larger radius will be important, since the
actual electron fraction does track its equilibrium value so closely.
As the $\nu_e$'s convert to sterile species, the actual electron
fraction falls very quickly through 1/3, and $\bar{\nu}_e$'s begin to
convert to steriles.  Both because $Y_e$ is falling so quickly and
because this first $\bar{\nu}_e$ resonance occurs at a position where
there is a comparatively large density gradient, the neutrino
amplitude evolution through the resonance may not be completely
adiabatic.  The second $\bar{\nu}_e$ resonance, however, will be
adiabatic as usual.  Thus, some $\bar{\nu}_e$'s will not convert in
the first resonance, but will convert to steriles in the second.  The
net effect is to lower the population of $\bar{\nu}_e$'s, and
therefore raise the equilibrium value of $Y_e$.  This effect is
minimized if the resonances occur far from the surface of the neutron
star, where they are largely adiabatic.

In both the $\tau = 0.1$ s and $\tau = 0.9$ s cases, the minimum in
final $Y_e$ is larger than the minimum in the $\tau = 0.3$ s case.  In
the limit of very short dynamic expansion timescale, the number of
neutrino captures after the $\nu_e$ resonance is very small and the
electron fraction remains high.  For example at $\tau=0.01$ s,
neutrino conversion has very little effect on the electron fraction.
The region of low $Y_e$ will disappear at very low expansion rate,
owing to the strengthening of the alpha effect in this limit.

An alternative solution to the $r$-process problem would be to invoke
a very rapid outflow in the absence of neutrino flavor transformation.
This suppresses both the alpha effect and the assembly of seed nuclei,
therefore increasing the neutron-to-seed ratio.  However, it is not
obvious that the neutrino heating mechanism can be responsible for
such rapid ejection.

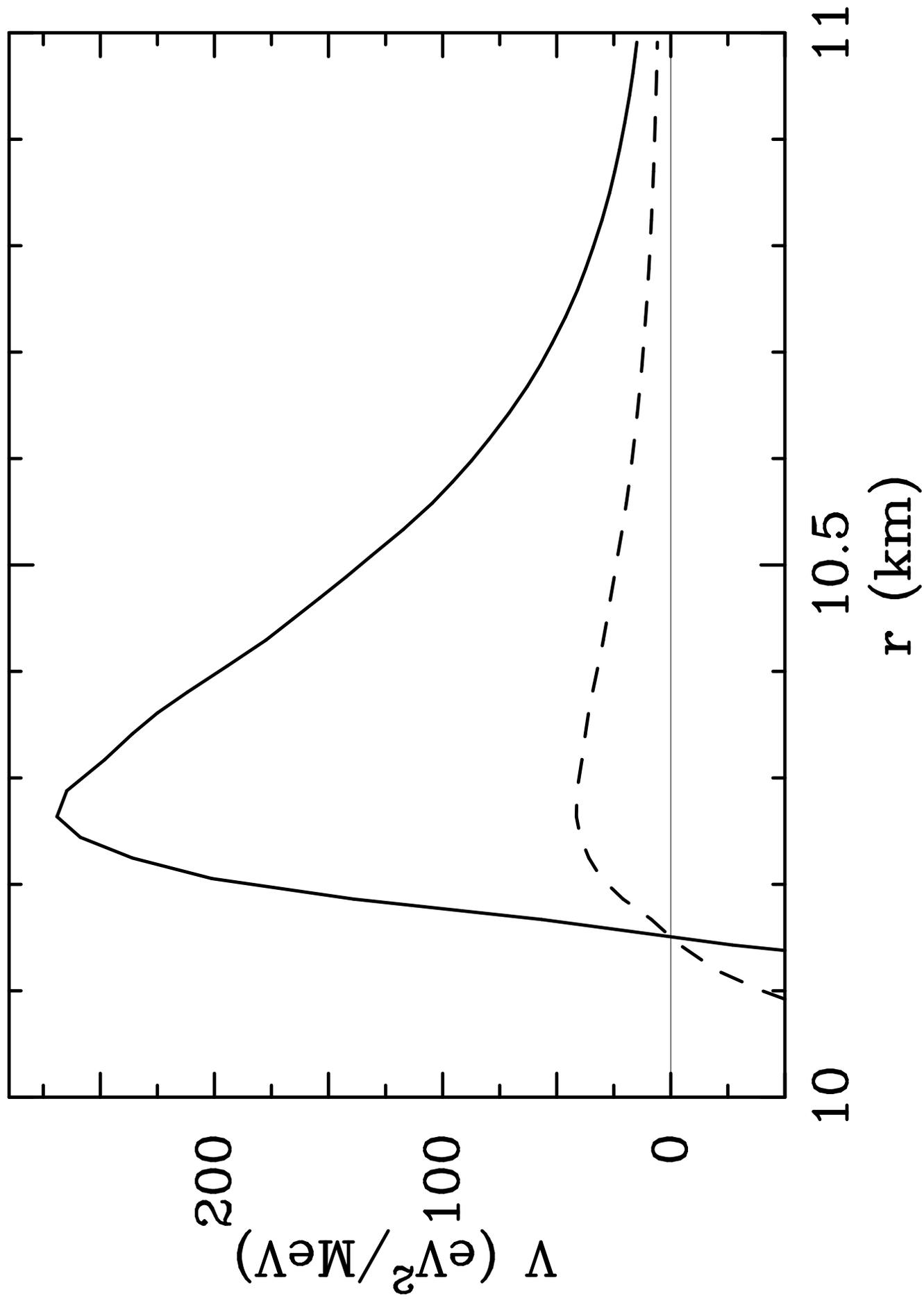
\begin{figure}
\caption{The potentials, $V(r) \propto \rho(Y_e - 1/3)$, for the two
  density profiles which we consider. The solid line shows the
  potential we used for most of our calculations; the dashed line
  shows the potential for the alternate density profile (see Section
  IV.B).}
\label{fig:scale} \end{figure}

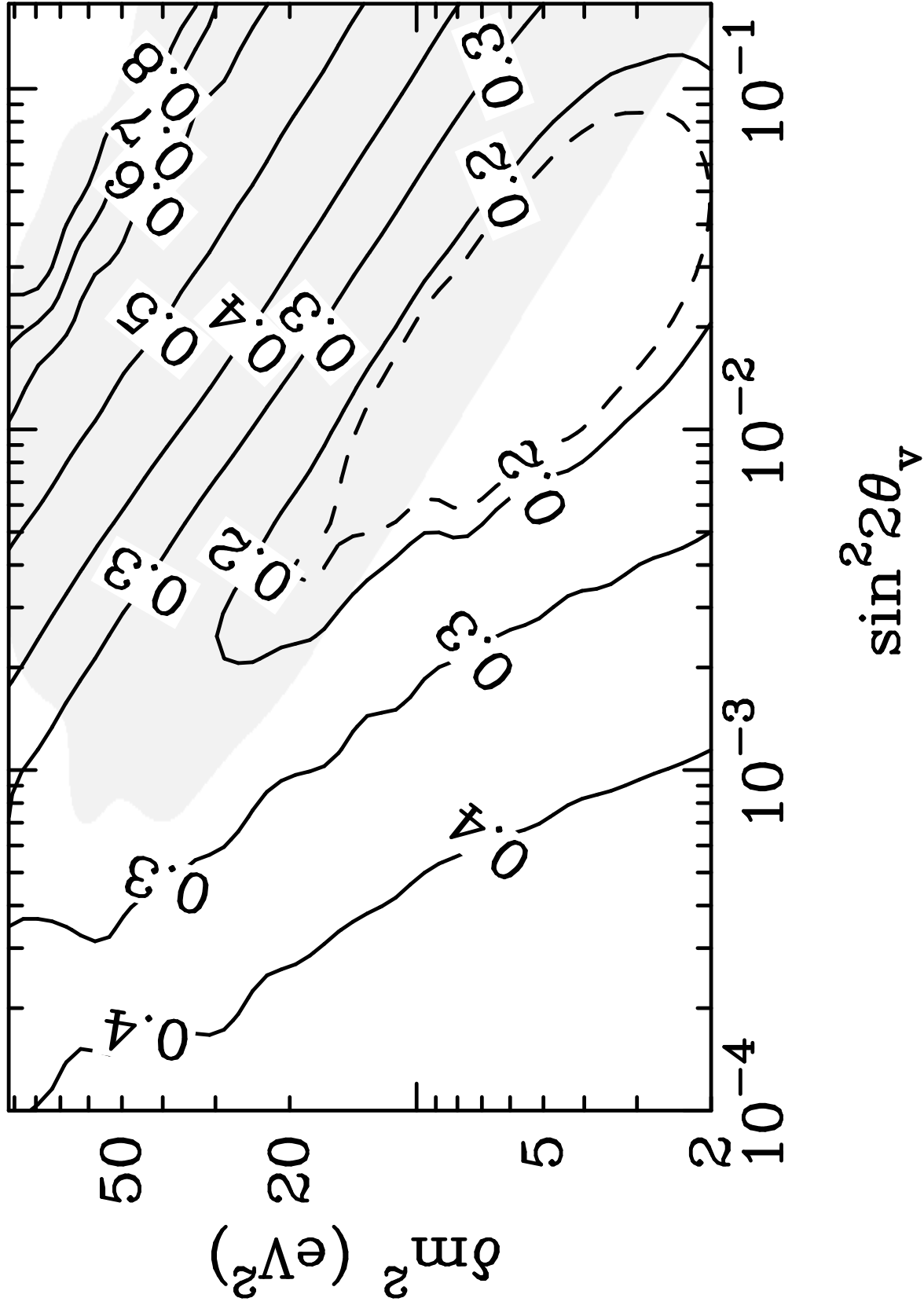
\begin{figure} 
\caption{Contour plot generated with alternate density profile.
  In the grey region nonradial $\nu$ paths may be
  significant. Nonradial paths will not be important at high $\delta
  m^2$ because the potential $V(r)$ turns over at a fairly small
  value.}  
\label{fig:sc-cont} 
\end{figure}

In addition to the variation of parameters in the wind model, one must
also consider variation in the density profile before the wind takes
over.  This is particularly important since if this is less steep than
in our example, there will be more conversion of electron neutrinos in
the first resonance, potentially destroying the low $Y_e$ solution
that we have presented.  We tested this by employing a different,
unrealistically flat density gradient in the intermediate region, the
potential for which is shown in Figure \ref{fig:scale}, and generating
the same type of contour plot as in our main example.  The results are
shown in Figure \ref{fig:sc-cont}.  Clearly, this part of the density
profile has a quantitative impact on the solution, although it does
not change our qualitative conclusions.

\section{Conclusions}

Here we have followed in the region above a hot protoneutron star the
evolution of the $\nu_e$ and $\bar\nu_e$ neutrino distribution
functions including active-sterile neutrino transformation in the
channels $\nu_e \rightleftharpoons \nu_s$ and $\bar\nu_e
\rightleftharpoons \bar\nu_s$. This evolution was calculated from the
surface of the neutron star through the region in which the key input
quantities for $r$-process nucleosynthesis are determined. We employed
a realistic outflow model which included feedback effects from
material expansion and neutrino flavor/type evolution and which
included a nuclear equation of state sophisticated enough to model
adequately the alpha effect.

We have found that a very interesting range of vacuum neutrino
mass-squared differences $3\,{\rm eV^2} {\ 
  \lower-1.2pt\vbox{\hbox{\rlap{$<$}\lower5pt\vbox{\hbox{$\sim$}}}}\ }
\delta m^2_{\rm es} {\ 
  \lower-1.2pt\vbox{\hbox{\rlap{$<$}\lower5pt\vbox{\hbox{$\sim$}}}}\ q
  }
  70\,{\rm eV^2}$ and vacuum mixing angles $\sin^22\theta_{\rm es} {\ 
    \lower-1.2pt\vbox{\hbox{\rlap{$>$}\lower5pt\vbox{\hbox{$\sim$}}}}\ 
    } {10}^{-3}$ produces effects which favor an increase in the
  neutron to seed nucleus ratio.  (Here, $\delta m^2_{\rm es}$ and
  $\theta_{\rm es}$ refer to the parameters that control the $\nu_e
  \rightleftharpoons \nu_s$ and $\bar\nu_e \rightleftharpoons
  \bar\nu_s$ evolution.)  In fact, the optimal range in neutrino
  mixing parameters produces a greatly reduced electron fraction $Y_e$
  and a significantly smaller population of $\nu_e$'s irradiating the
  nucleosynthesis region. These effects act to aid $r$-process
  nucleosynthesis in two ways: (1) the lower $Y_e$ translates directly
  into more neutrons that can be captured to make the heavy
  $r$-process nuclides; and (2), the diminished flux of $\nu_e$'s
  helps to disable the pernicious alpha effect, which is a serious
  obstacle to obtaining $r$-process nucleosynthesis in neutrino-heated
  supernova ejecta.

These effects that are beneficial to the $r$-process in this site
come about through the disproportionate disappearance of the $\nu_e$
population relative to the $\bar\nu_e$'s. In turn, the reason that
so many more $\nu_e$'s are converted to sterile species than
$\bar\nu_e \rightarrow \bar\nu_s$ in our calculation has to do with
a new effect which we point out here for the first time. A self
consistent calculation of the electron fraction $Y_e$ with neutrino
transformation and with a proper treatment of the material outflow
rate shows that although the $\bar\nu_e$ are converted to sterile
species, they are re-generated before the $Y_e$ in the wind freezes
out. This behavior also prevents the system from reaching the fixed
point in its evolution, $Y_e =1/3$.

A proper treatment of expansion coupled with $Y_e$ evolution is a
necessary step in obtaining this new result. In fact, we find that
the weak potential driving $\bar\nu_e \rightleftharpoons \bar\nu_s$
immediately past the radius where $Y_e$ crosses below $1/3$ has a
peaked structure with radius. This potential at first rises as the
quantity $\left(1/3 - Y_e \right)$ rises, and then falls with
increasing radius as the baryon density drops. This implies that
there will be two level crossings (resonances) in the $\bar\nu_e
\rightleftharpoons \bar\nu_s$ channel within a short space in
radius.  Therefore $\bar\nu_e$'s converted at the first resonance
are re-generated at the second.

We have employed several approximations in obtaining this result, some
of which we are pursuing with further investigation.  For convenience
in computation, we have treated the neutrino flux as arising from a
point source.  This is unphysical on several grounds: it leads to a
too rapid fall-off with radius of neutrino and antineutrino capture
rates; and it implies no neutrino background terms in the weak
potential. As outlined above, however, we expect the neutrino
background to dominate the weak potential where $Y_e \approx 1/3$.
Just where and to what extent the background will dominate and alter
the neutrino flavor evolution from what we have presented here depends
on the neutrino energy spectra and luminosities.  In turn, there
exists a wide range of possible values for these quantities at the
late epoch where $r$-process nucleosynthesis is an issue. This
variation in neutrino emission parameters reflects the range of
nuclear equations of state, neutrino opacities, and neutrino transport
physics employed in the various numerical computations. We are
investigating the ranges of late-time supernova conditions for which
our effect will be operative.

We have implicitly assumed here that the de-leptonizing neutron star
is not itself a source of sterile neutrinos. This seems reasonable on
two grounds.  First, we can invoke very small vacuum mixing angles
between active and sterile neutrino species. That suits our purposes
in the late-time neutrino-driven wind because we rely on
matter-enhancement of neutrino flavor transformation. For the neutrino
mass differences employed in this paper there will be no mass level
crossings deep in the core and, in fact, matter effects will then
further suppress mixing. Second, the neutrino mean free paths in the
core can be very short, so that coherent flavor transformation is
unlikely. The problem of neutrino production, interaction, and
propagation in dense and hot nuclear matter is a difficult one and
merits much further study.

Of course, it is never legitimate to invoke novel weak interaction
physics at some point late in the evolution of the supernova without
an assessment of how this new physics could have altered the picture
in earlier epochs. In particular, what would be the effects of the
neutrino mass and mixing scheme we invoke here to help the $r$-process
on the core infall epoch, and on the shock re-heating epoch? The
infall of the pre-supernova iron core is characterized by low entropy
per baryon, relatively high densities ($\rho {\ 
  \lower-1.2pt\vbox{\hbox{\rlap{$>$}\lower5pt\vbox{\hbox{$\sim$}}}}\ }
{10}^{11}\,{\rm g {cm}^{-3}}$ where neutrinos are trapped and at least
partially thermalized) and, with the typical equations of state
employed, electron fractions $Y_e > 1/3$. These conditions imply that
our chosen range of neutrino mass and mixing parameters will produce
no neutrino mass level crossings which could alter the standard core
collapse picture. In short, the required $\delta m^2$ to obtain a
level crossing on infall is much larger (see Ref. \cite{FMWS87}).

Likewise, the $\delta m^2$ values required to obtain a neutrino mass
level crossing under the shock during the supernova explosion (or
shock re-heating) epoch are large $\delta m^2 {\ 
  \lower-1.2pt\vbox{\hbox{\rlap{$>$}\lower5pt\vbox{\hbox{$\sim$}}}}\ 
  }100\,{\rm eV}^2$ (see Ref. \cite{FMMW92} for a discussion of
active-active neutrino flavor transformation during shock re-heating).
At this epoch, however, we are much less certain about the range of
$Y_e$ values likely to be encountered either near the surface of the
core or in the higher entropy material behind the shock.

In fact, we expect this epoch to be accompanied by salt-finger-like
convective instability through the neutrino sphere, which could
greatly increase the neutrino luminosities and neutrino heating rates
immediately above the hot proto-neutron star surface. In turn, this
increased heating probably leads to convection and and to large and
small scale inhomogeneities in density, entropy, and electron
fraction.  It is possible that the fluctuation amplitudes on relevant
scales at this epoch will be large enough to destroy complete
adiabatic neutrino flavor evolution through resonances and, hence,
render our scheme inoperative at these early times. Note, however,
that it may be reasonable to assume that the material outflow is much
smoother and so conducive to adiabatic neutrino flavor evolution in
the later neutrino-driven outflow regime where we envision the
$r$-process to originate and where our scheme could operate.
Fluctuation-induced neutrino flavor de-polarization in the context of
supernovae and the sun has been investigated in detail in the
active-active channel \cite{fluct}, and these studies should be
directly applicable as well in the active-sterile channel employed
here.

We intend to investigate the effects of active-sterile mixing schemes
on shock re-heating and on the production of the neutron number N$=50$
nuclei and the light p-nuclei which may originate during this epoch
\cite{alpha,HWFM96}. In particular, if large scale neutrino flavor
transformation somehow {\it does} occur at this epoch, then a
reduction in the electron fraction would exacerbate the existing
problem in some supernova models of the overproduction of N$=50$
nuclei. Ultimately, since the conditions during the shock re-heating
epoch and the later neutrino-driven wind epoch are so disparate, we
feel that our new scheme to help the $r$-process stands on its own.

Our results are potentially significant in the debate over {\it where}
$r$-process nucleosynthesis takes place in the Galaxy. There is fair
evidence that at least some of the $r$-process nuclides are made in an
environment associated with core collapse supernovae (Type II, Type Ib
and Ic supernovae) \cite{QVW98}.

As outlined above, it is so far difficult to obtain conditions
favorable for $r$-process nucleosynthesis in conventional
neutrino-driven outflow models\cite{MMF}, especially when the alpha
effect is included \cite{alpha}. It is an open question as to whether
or not the problems with $r$-process nucleosynthesis in this site can
be remedied through the tuning of the astrophysical aspects of the
outflow model.  Furthermore, it is not really known whether it is {\it
  required} to have $r$-process nucleosynthesis come from this site in
order to explain the observational and meteoritic data (see
Ref.~\cite{Fuller}). It has been argued, however, that neutrino
post-processing may be important in understanding the observed
abundance patterns and this may imply that supernovae or neutron star
binary mergers, or both, play a role in $r$-process synthesis
\cite{haxton}.

Based on our work here we can say, however, that if neutrino mass and
mixing parameters are in our optimal range, then a broad class of
neutrino-driven outflow models have the {\it necessary} conditions to
produce the $r$-process.  Moreover, we would obtain the $r$-process in
these models in a way which was robust to the details and
astrophysical uncertainties in the models over a fairly broad range of
outflow parameters. The issue of {\it sufficiency} of $r$-process
nucleosynthesis in this case is another matter and could only be
answered with a detailed nuclear reaction network which included
neutron capture and photo-disintegration all coupled with a consistent
hydrodynamic calculation, as well as all of our neutrino physics
effects.

So, does the existence of $r$-process nuclides in the abundances
measured in the Sun and other stars and with the synthesis rates
inferred in the Galaxy then imply the existence of light sterile
neutrinos? The answer is no, since we cannot at this time preclude
other non-neutrino-mixing astrophysical fixes for the $r$-process, and
we cannot say for absolutely certain that we need the $r$-process from
neutrino-heated supernova ejecta. Nevertheless, it is interesting that
light sterile neutrinos mixing with electron neutrinos could affect
the synthesis of the heaviest elements. At the present time there is a
flurry of new instruments which are bringing in new data which bears
on the issues surrounding $r$-process nucleosynthesis, so it may be
possible in the future to resolve uncertainties. If there truly are
\lq\lq sterile\rq\rq\ neutrinos, then astrophysical means, principally
nucleosynthetic, represent probably our only hope for learning about
their properties.

Finally, what of the implications of our results for particle physics?
In our models we ignore the mu and tau neutrinos and their
anti-particles as these play a negligible role in the nucleosynthesis
scenario considered here.  Our model would, for example, be consistent
with having the sterile neutrino mass at around $\sim 2 - 8\,{\rm eV}$
while having all of the active neutrinos clustered near zero mass,
with the $\nu_e$ and $\nu_\mu$/$\nu_\tau$ split by $\sim
{10}^{-10}\,{\rm eV^2}$ to $\sim {10}^{-4}\,{\rm eV^2}$ to give the
favored solar neutrino solutions, and to have the mu and tau neutrino
maximally mixed with their masses split by some $\sim {10}^{-2}\,{\rm
  eV^2}$ to give the Superkamiokande result for atmospheric neutrinos.
The LSND result could be accommodated in our model by invoking an
indirect vacuum oscillation of $\bar\nu_\mu$ into $\bar\nu_e$ via the
sterile species, $\bar\nu_\mu \rightarrow \bar\nu_s \rightarrow
\bar\nu_e$ \cite{bffm}. By contrast, other schemes involving sterile
neutrinos designed to fix the $r$-process \cite{CFQ98}, would predict
active-sterile mixing in the sun as a solution of the solar neutrino
problem.  We may have a resolution of this question from the Sudbury
Neutrino Observatory in the near future. In any case, the future is
promising for the role of $r$-process studies to help constrain
neutrino mass and mixing models. In turn, future neutrino oscillation
experiments conceivably could help us to constrain the site of
$r$-process nucleosynthesis.


\section*{ACKNOWLEDGMENTS}

This work was supported in part by the U.S. National Science
Foundation Grants No.\ PHY-9605140 at the University of Wisconsin, and
PHY-9800980 at the University of California, San Diego, and in part by
the University of Wisconsin Research Committee with funds granted by
the Wisconsin Alumni Research Foundation. We thank the Institute for
Nuclear Theory and Department of Astronomy at the University of
Washington, and Aspen Center for Physics for their hospitality and
Department of Energy for partial support during the completion of this
work.


\newpage

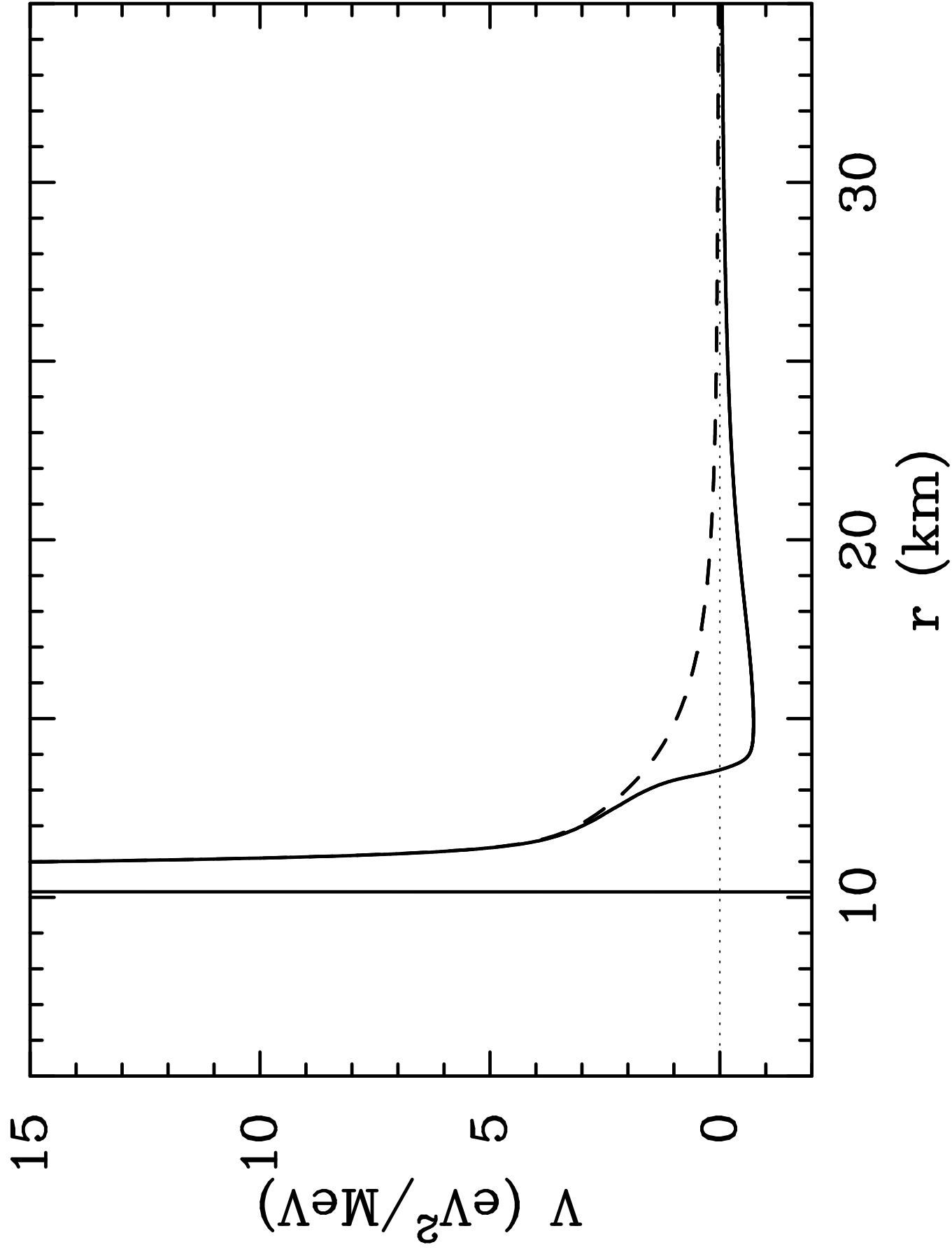
\begin{figure}
\caption{The potential, $V$, is plotted against distance (solid
line).  For comparison we also show $V$ when feedback effects are
not included (dotted line).  The nearly vertical line at the left edge
of the plot corresponds to the inmost $\bar{\nu}_e$ and $\nu_e$
resonance at the surface of the neutron star.}
\label{fig:potl}
\end{figure}
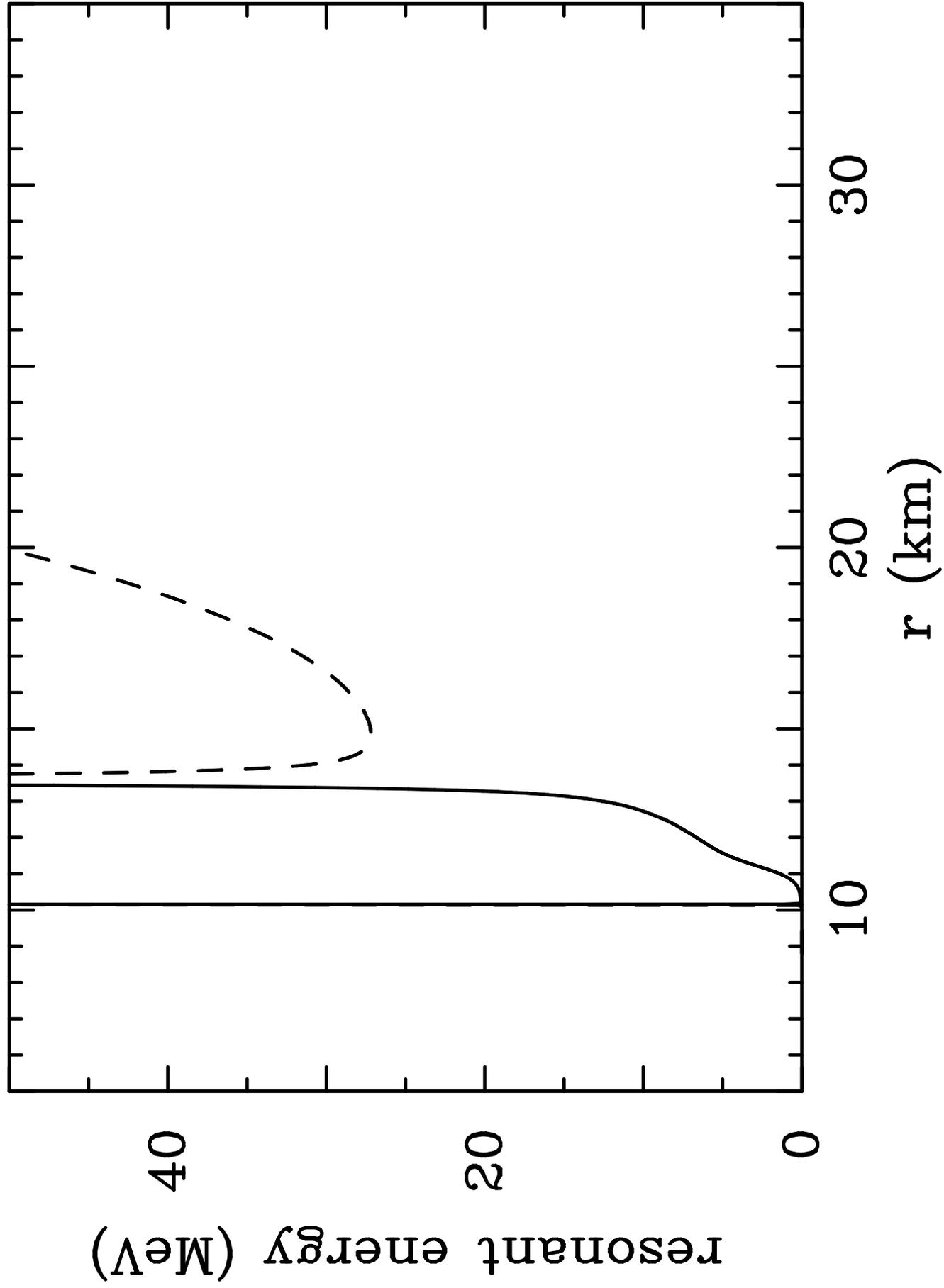
\begin{figure}

\epsfxsize=14cm
\epsfxsize=14cm
\caption{Energy of $\nu_e$s (solid line) and $\bar{\nu}_e$s (dashed
line) undergoing resonance plotted against distance from the center of
the protoneutron star, for $\sin^2 2 \theta_{\rm v} = 0.01$, and
$\delta m^2 = 20 \, {\rm eV}^2$, and a dynamical time of $\tau = 0.3
\, {\rm s}$.  For this choice of parameters, electron antineutrinos
below $\sim 25$ MeV never undergo a resonance beyond the surface of
the neutron star.  A resonance may cause a near complete or partial
flavor transformation, depending on the adiabaticity.}
\label{fig:energy}
\end{figure}

\newpage
\begin{references}

\bibitem{bbfh}
  E.M. Burbidge, G.R. Burbidge, W.A. Fowler, and F. Hoyle, Rev. Mod.
  Phys. {\bf 29}, 547 (1957); A. G. W. Cameron,
  Proc.~Astron.~Soc.~Pacific {\bf 69}, 201 (1957).
\bibitem{QVW98}  
  Y.-Z. Qian, P. Vogel, and G. J. Wasserburg, Astrophys. J. {\bf 494},
  285 (1998).
\bibitem{neudrw}
   S.E. Woosley and R.D. Hoffman, Astrophys. J. {\bf 395}, 202 (1992).
\bibitem{r1}   
  S.E. Woosley, J.R. Wilson, G.J. Mathews, R.D. Hoffman, and B.S.
  Meyer, Astrophys. J. {\bf 433}, 229 (1994); B.S. Meyer, W.M. Howard,
  G.J. Mathews, S.E. Woosley, and R.D. Hoffman, Astrophys. J. {\bf
    399}, 656 (1992).  \bibitem{r2} K. Takahashi,J.  Witti, and H.-Th.
  Janka, Astron. Astrophys. {\bf 286}, 857 (1994).
\bibitem{meyeralpha} 
  B.S. Meyer, Astrophys. J., {\bf 449}, L55 (1995).
\bibitem{qw} 
  Y.-Z. Qian and S.E. Woosley, Astrophys. J., {\bf 471},
  331 (1996).  
\bibitem{wilson} 
  J.R. Wilson, as quoted in Ref.~\cite{r1}.  
\bibitem{putthisin1} 
  R.  D. Hoffman, S. E. Woosley, and Y.-Z. Qian, Astrophys. J. {\bf
    482}, 951 (1996).
\bibitem{meyer97} 
  B. S. Meyer and J. S. Brown, Astrophys. J. Suppl.  {\bf 112}, 199
  (1997).
\bibitem{CF} 
  C. Y.  Cardall and G. M. Fuller, Astrophys. J. Lett.  {\bf 486},
  L111 (1997).
\bibitem{FQ96} 
  G. M.  Fuller and Y.-Z. Qian, Nucl. Phys. A {\bf 606}, 167 (1996).
\bibitem{alpha} 
  G.M. Fuller and B.S. Meyer, Astrophys. J., {\bf 453}, 792 (1995).
\bibitem{MMF} 
  B. S. Meyer, G.  C. McLaughlin, and G. M.  Fuller, Phys. Rev. C,
  {\sl in press} (1998).
\bibitem{Qetal} 
  Y.-Z.  Qian, {\em et el.}, Phys. Rev. Lett.  {\bf 71}, 1965 (1993).
\bibitem{QF95} 
  Y.-Z. Qian and G.M. Fuller, Phys. Rev. D {\bf 51}, 1479 (1995), {\bf
    52}, 656 (1995).
\bibitem{Sigl} 
  G. Sigl, Phys. Rev. D {\bf 51}, 4035 (1995).
\bibitem{Pelt} 
  J. T. Peltoniemi, Astron.  Astrophys., {\bf 254}, 121 (1992); J. T.
  Peltoniemi, preprint hep-ph/9511323, unpublished (1995).
\bibitem{CFQ98} 
  D. O. Caldwell, G. M. Fuller, and Y.-Z.  Qian, Phys. Rev. D, {\sl
    submitted} (1998).
\bibitem{solar} 
  Y.  Fukuda, {\em et al.}, (The Superkamiokande collaboration), Phys.
  Rev. Lett. {\bf 81}, 1158 (1998).  \bibitem{othersolar} B.T.
  Cleveland, T. Daily, R. Davis, J.R. Distel, K. Lande, C.K. Lee, P.S.
  Wildenhain, and J. Ullman, Astrophys. J.  {\bf 496}, 505 (1998); W.
  Hampel, {\em et al.} (GALLEX Collaboration), Phys. Lett. B {\bf
    388}, 384 (1996), {\bf 420}, 114 (1998), {\em ibid.} {\bf 436},
  158 (1998); J.N. Abdurashitov, {\em et al.} (SAGE Collaboration),
  Nucl.  Phys. Proc. Suppl. {\bf 48}, 299 (1996), hep-ph/9803418.
\bibitem{solaran}   
  J.N. Bahcall, P.I.  Krastev, and A.Yu. Smirnov, Phys.  Rev. D {\bf
    58}, 096016 (1998); V. Barger, S. Pakvasa, T. J. Weiler, and K.
  Whisnant, {\em ibid.}, 093016 (1998); N. Hata and P.  Langacker,
  {\em ibid} {\bf 56}, 6107 (1997).
\bibitem{atm} 
  Y. Fukuda, {\em et al.}, (The Superkamiokande collaboration), Phys.
  Rev. Lett. {\bf 81}, 1562 (1998); Phys. Lett. B {\bf 436}, 33
  (1998). 
\bibitem{chooz} 
  M. Apollonio, {\em et al.} (The CHOOZ collaboration), Phys. Lett. B
  {\bf 420}, 397 (1998).  
\bibitem{atman} 
  M. Fukugita, M. Tanimoto, and T. Yanagida, Phys. Rev.  D {\bf 57},
  4429 (1998); R. Foot, R. R. Volkas, and O.  Yasuda, Phys. Rev. D
  {\bf 58}, 013006 (1998); Q.Y. Liu and A.Yu.  Smirnov, Nucl. Phys. B
  {\bf 524}, 505 (1998); G.L. Fogli, E. Lisi, A.  Marrone, and D.
  Montanino, Phys. Lett. B {\bf 425}, 341 (1998); M.  C.
  Gonzalez-Garcia, H. Nunokawa, O. L. G. Peres, T. Stanev, and J.  W.
  F. Valle, Phys. Rev. D {\bf 58}, 033004 (1998); P. Lipari and M.
  Lusignoli, Phys. Rev. D {\bf 58} 073005 (1998); M. C.
  Gonzalez-Garcia, H. Nunokawa, O. L. G. Peres, and J. W. F.  Valle,
  hep-ph/9807305.  V. Barger, T. J. Weiler, and K. Whisnant, Phys.
  Lett. B {\bf 440}, 1 (1998); G.L. Fogli, E. Lisi, A. Marrone, and G.
  Scioscia, Phys. Rev. D {\bf 59}, 033001 (1999).
\bibitem{lsnd} 
  C. Athanassopoulos, {\em et al.}  (LSND Collaboration), Phys. Rev.
  Lett. {\bf 75}, 2650 (1996), {\bf 77}, 3082 (1996), Phys. Rev. C
  {\bf 54}, 2685 (1996), Phys. Rev. C {\bf 58}, 2489 (1998); see also
  J.E. Hill, Phys. Rev. Lett. {\bf 75}, 2654 (1996).
\bibitem{3nus} 
  C. Y. Cardall and G. M. Fuller, Phys. Rev. D{\bf 53}, 4421 (1996);
  G.  L. Fogli, E. Lisi, and G.  Scioscia, Phys. Rev. D{\bf 52}, 5334
  (1995); A. Acker and S.  Pakvasa, Phys. Lett. {\bf B397}, 209
  (1997); P. Harrison, D.  Perkins, and W.Scott, Phys. Rev. {\bf
    B349}, 137 (1995); R. P. Thun and S. Mckee, Phys. Lett. B {\bf
    439}, 123 (1998); V.  Barger, T. J. Weiler, and K.  Whisnant,
  Phys. Lett. B {\bf 440}, 1 (1998); T.  Teshima and T. Sakai
  hep-ph/9805386 and hep-ph/9801276.
\bibitem{cald} 
  D. O. Caldwell and R. N. Mohapatra, Phys. Rev. D{\bf 48}, 3259
  (1993); D. O.  Caldwell and R. N.  Mohapatra, Phys. Lett. B {\bf
    354}, 371 (1995); G.  M.  Fuller, J.  R. Primack, and Y.-Z. Qian,
  Phys. Rev. D{\bf 52}, 1288 (1995).
\bibitem{msw} 
  S.P. Mikheyev and A. Yu. Smirnov, Sov. J. Nucl. Phys.  {\bf 42}, 913
  (1985); Sov. Phys. JETP {\bf 64}, 4 (1986); L.  Wolfenstein, Phys.
  Rev. D {\bf 17}, 2369 (1978); {\em ibid.\/} {\bf 20}, 2634 (1979).
\bibitem{nunokawa}  
  H. Nunokawa, J. T.  Peltoniemi, A. Rossi, and J. W. F. Valle, Phys.
  Rev. D {\bf 56}, 1704 (1997).
\bibitem{hax-lz} 
  W. Haxton, Phys. Rev. Lett. {\bf 57}, 1271 (1986); S.J. Parke, {\em
    ibid.}, 1275 (1986).
\bibitem{bb-lz}
  H.A. Bethe, Phys. Rev. Lett. {\bf 56}, 1305 (1986).  
\bibitem{sigl}
  G. Sigl and G. Raffelt, Nucl. Phys. B {\bf 406}, 423 (1993).
\bibitem{wilsonmayle} 
  R. W. Mayle and J. R. Wilson, unpublished (1993).
\bibitem{MFW} 
  G. C. McLaughlin, G. M. Fuller, and J. R.  Wilson, Astrophys. J.
  {\bf 472}, 440 (1996).
\bibitem{FMWS87} 
  G. M.  Fuller, R. Mayle, J. R. Wilson, and D. N. Schramm, Astrophys.
  J., {\bf 322}, 795 (1987).
\bibitem{FMMW92} 
  G. M. Fuller, R. Mayle, B.  S. Meyer, and J. R. Wilson, Astrophys.
  J., {\bf 389}, 517 (1992).
\bibitem{fluct}   
  R.F. Sawyer, Phys. Rev. D {\bf 42}, 3908 (1990); F.N.  Loreti, Y.Z.
  Qian, G.M. Fuller, and A.B. Balantekin, Phys. Rev. D {\bf 52}, 6664
  (1995). 
\bibitem{HWFM96}   
  R. Hoffman, S. E. Woosley, G. M. Fuller, and B. S. Meyer, Astrophys.
  J., {\bf 460}, 478 (1996).
\bibitem{Fuller} 
  G. M. Fuller, AIP Conference Proceedings {\bf 412}, 160 (1997).
\bibitem{MKC}
  Meyer, B. S., T. D. Krishnan, and D.D. Clayton, {\bf 498}, 808
  (1998).
\bibitem{haxton} 
  W.C. Haxton, K. Langanke, Y.Z. Qian, and P. Vogel, Phys. Rev. Lett.
  {\bf 78}, 2694 (1997).
\bibitem{bffm} 
  A. B. Balantekin, J. Fetter, G. M. Fuller, and G. C.  McLaughlin,
  Phys. Rev. Lett., to be submitted (1998).

\end{references}
\end{document}